# Global regolith thermophysical properties of the Moon from the Diviner Lunar Radiometer Experiment

Paul O. Hayne[1,*], Joshua L. Bandfield[2], Matthew A. Siegler[3], Ashwin R. Vasavada[1], Rebecca R. Ghent[3,4], Jean-Pierre Williams[5], Benjamin T. Greenhagen[6], Oded Aharonson[7], Catherine M. Elder[1], Paul G. Lucey[8], and David A. Paige[5]

[1]Jet Propulsion Laboratory, California Institute of Technology, USA
[2]Space Science Institute, USA
[3]Planetary Science Institute, USA
[4]Department of Earth Sciences, University of Toronto, Canada
[5]Deparment of Earth, Planetary and Space Sciences, University of California, Los Angeles, USA
[6]Applied Physics Laboratory, Johns Hopkins University, USA
[7]Department of Earth and Planetary Sciences, Weizmann Institute of Science, Rehovot, Israel
[8]University of Hawaii, Manoa, USA

[*]**Corresponding author**. **Address:** Jet Propulsion Laboratory, California Institute of Technology, MS 183-301, 4800 Oak Grove Drive, Pasadena, CA 91109, USA. **E-mail address:** Paul.O.Hayne@jpl.nasa.gov

**Highlights:**

- We present global maps of lunar regolith thermophysical properties
- The Moon's upper ~5 cm of regolith has a globally averaged thermal inertia of ~55 ±2 J m$^{-2}$ K$^{-1}$ s$^{-1/2}$ at a reference temperature of 273 K
- Variations in regolith thermophysical properties on regional and local scales are due primarily to impact processes < 1 Ga
- Thermophysical properties of ejecta blankets can be used as a chronometer for impact craters of age ~1 Ma to 1 Ga
- Regional scale pyroclastic deposits have low thermal inertia, constraining their eruption styles and thickness

**Abstract:** We used infrared data from the Lunar Reconnaissance Orbiter (LRO) Diviner Lunar Radiometer Experiment to globally map thermophysical properties of the Moon's regolith fines layer. Thermal conductivity varies from 7.4×10$^{-4}$ W m$^{-1}$ K$^{-1}$ at the surface, to 3.4×10$^{-3}$ W m$^{-1}$ K$^{-1}$ at depths of ~1 m, given density values of 1100 kg m-3 at the surface, to 1800 kg m$^{-3}$ at 1-m depth. On average, the scale height of these profiles is ~7 cm, corresponding to a thermal inertia of 55 ±2 J m$^{-2}$ K$^{-1}$ s$^{-1/2}$ at 273 K, relevant to the diurnally active near-surface layer, ~4-7 cm. The temperature-dependence of thermal conductivity and heat capacity leads to a ~2x diurnal variation in thermal inertia at the equator. On global scales, the regolith fines are remarkably uniform, implying rapid homogenization by impact gardening of this layer on timescales < 1 Gyr. Regional and local scale variations show prominent impact features < 1 Gyr old, including higher thermal inertia (> 100 J m$^{-2}$ K$^{-1}$ s$^{-1/2}$) in the interiors and ejecta of Copernican-aged impact craters, and lower thermal inertia (< 50 J m$^{-2}$ K$^{-1}$ s$^{-1/2}$) within the lunar cold spots identified by Bandfield et al. (2014). Observed trends in ejecta thermal inertia provide a potential tool for age-dating craters of previously unknown age, complementary to the approach suggested by Ghent et al. (2014). Several anomalous regions are identified in the global 128 pixels-per-degree maps presented here, including a high-thermal inertia deposit near the antipode of Tycho crater.

# 1 Introduction

Regolith is the layer of loose material covering the surfaces of the Moon and many other solar system bodies. This porous, granular layer thermally insulates underlying bedrock and absorbs cosmic rays. Regolith also records the history of fragmentation and overturn by hypervelocity meteoroid impacts, which dominate the last billion years of lunar geologic history (Shoemaker et al., 1969; 1970; Oberbeck et al., 1973; Melosh, 1989; McKay et al., 1991). Over much of the Moon, the primary regolith has a characteristic thickness of several meters (Shkuratov and Bondarenko, 2001; Fa and Wieczorek, 2012). Because the impact flux is dominated by the smallest impactors, the upper layers of the lunar surface are overturned and pulverized much more frequently than lower layers (Gault, 1974; Arnold 1975). Apollo core samples showed depth-dependent density and thermal conductivity profiles, presumably caused by this vertical variation in overturn timescale (Mitchell et al., 1973) in addition to compaction by overburden stress. Therefore, the uppermost regolith is composed of finer-grained, highly porous material characterized by low bulk density and thermal conductivity. Density and conductivity increase with depth, as does the prevalence of larger rock fragments. These properties strongly affect surface and subsurface temperatures. Local and regional differences in regolith properties may also reveal overturn histories important for understanding cosmic ray exposure ages of individual samples (Langevin and Arnold, 1977).

Previous investigations using in-situ and remote observations have revealed both subsurface and lateral variations in regolith properties. Core samples acquired by the astronauts at the Apollo landing sites showed a general increase in density with depth, but with significant stratification in both composition and thermophysical properties due to overlapping impact ejecta (Carrier et al., 1991). Nonetheless, remote sensing and in-situ temperature measurements can be well matched by modeling the subsurface density and conductivity profiles as continuous functions, perhaps with a discrete upper porous layer ~1-2 cm thick (Keihm and Langseth, 1973b). This suggests that layering within the regolith is more or less spatially random on the scale of thermal diffusion: $z_s \sim \sqrt{\kappa P/\pi}$, where $\kappa$ is the thermal diffusivity (~$10^{-9}$ m$^2$ s$^{-1}$), $P$ is the lunar synodic rotation period (~29.53 Earth days). The quantity $z_s$ is known as the thermal skin depth, and is ~4 – 10 cm for typical upper lunar regolith properties, or up to ~1 m for solid rock, on the diurnal timescale. Remote measurements of diurnal temperature cycles can therefore be used to constrain the thermophysical properties of the uppermost 10s of centimeters of lunar regolith.

Temperature cycles on the Moon have been measured using surface probes (Langseth et al., 1976), telescopic earth-based microwave and thermal infrared instruments (Pettit and Nicholson, 1930), and from lunar orbit (Paige et al., 2010a; Yu and Fa, 2016; Williams et al., 2017). Vasavada et al. (2012) studied the equatorial temperature cycles using the Diviner Lunar Radiometer instrument ("Diviner", described below) to derive various properties of the lunar surface materials. A key finding of that study was the high degree of spatial uniformity in these thermophysical properties. Subtle differences among geologic features were noted, but not studied in detail.

Here, we report a global investigation of the lunar regolith properties using Diviner radiometric measurements. In the following sections, we first describe the dataset and models used to constrain the lunar regolith properties. Results from this study show a variety of features (both expected and unexpected), revealing the rich geologic history of the Moon.

## 2  Dataset

Diviner is a nine-channel filter radiometer, with 2 solar channels and 7 infrared channels. Four of these infrared channels measure nighttime thermal emission: ~13-23, 25-41, 50-100, and 100-400 μm wavelengths (Paige et al., 2010a). Data from Diviner consist of brightness temperatures for each spectral channel ($T_b$) mapped at various local times on the lunar surface. Given the LRO spacecraft's polar orbit geometry, thermal maps are built up by sweeping each Diviner channel's 21-pixel array along a roughly north-south track, with a global mapping cycle completed in about one month (Williams et al., 2017). Since the start of the LRO science phase in mid-2009, Diviner has acquired data at multiple local times for virtually every location on the lunar surface. The resolution of the Diviner data is determined primarily by range to the surface; for a nominal spacecraft altitude of 50 km, the 3.6 by 6.1 mrad individual detector field-of-view provides a resolution of ~180 by 300 m on the lunar surface. Thus, for locations where coverage is nearly complete, gridded mapping can be performed at a resolution of 128 pixels-per-degree (ppd), which corresponds to ~250 m at the equator.

For the present study, we used the Level 2 gridded 128-ppd data products available on the Planetary Data System (Paige et al., 2011; Williams et al., 2016; 2017). In particular, we used derived rock abundance and regolith temperatures, which are described in detail by Bandfield et al. (2011; 2017). Their approach uses the spectral information from Diviner and modeled rock temperatures to discriminate between warm rocks and cooler regolith in the nighttime data. Bandfield et al. (2011, 2017) treated the rock abundance and rock-free regolith temperatures as free parameters that are allowed to vary within each pixel and local time. Rock abundance derived in this way represents the fractional area of exposed rocks large enough to be thermally isolated from the surrounding regolith, larger than ~10 – 100 cm in size (Section 3.3.1; cf. Bandfield et al., 2011). Thus, rocks smaller than ~10 cm augment the derived regolith temperatures and are considered part of the regolith in the study reported here. Although the abundance of exposed large rocks on the lunar surface is generally small (typically <1%), and therefore has little effect on the derived regolith temperatures and thermophysical properties (Appendix B), smaller rock fragments are an important component of the regolith thermophysical properties presented below. Similarly, buried rocks can influence surface temperatures if they are within ~10 cm of the surface (Elder et al., 2017a).

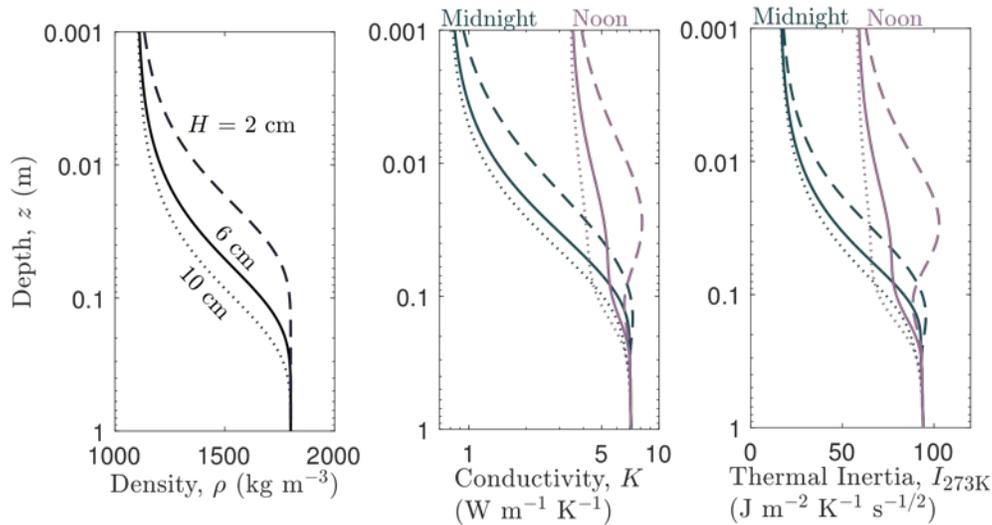

Figure 1: Regolith model profiles showing typical variations in density, thermal conductivity, and thermal inertia with depth. In this model, the upper and lower boundary conditions are fixed, and variations in the profiles are determined by the *H*-parameter. Three different values of the *H*-parameter are indicated: 2 cm (dashed curve), 6 cm (solid), 10 cm (dotted).

## 3   Approach and Methods

Here, we treat each location on the Moon as a discrete vertical column of regolith. Although unresolved heterogeneities may exist, thermophysical properties are derived for each 1/128th-degree (~250×250-m at the equator) pixel in aggregate. In the vertical dimension, regolith density and conductivity are assumed to increase downward, based on previous thermal studies and Apollo core samples (Carrier et al., 1973; Keihm and Langseth, 1973; Jones et al., 1975). Below we describe the retrieval approach adopted for deriving regolith profiles, along with its limitations.

### 3.1   Regolith Model

We use a one-dimensional thermal model (Appendix A) to derive thermophysical properties from the Diviner nighttime regolith temperatures. Daytime surface temperatures on the Moon are primarily controlled by surface energy balance through albedo and emissivity, whereas nighttime temperatures are controlled by thermophysical properties: thermal conductivity ($K$), density ($\rho$), and heat capacity ($c_p$). Based on model fits to equatorial brightness temperatures (Vasavada et al., 2012), the variation of density with depth $z$ can be accurately modeled as

$$\rho(z) = \rho_d - (\rho_d - \rho_s)e^{-\frac{z}{H}} \qquad (1)$$

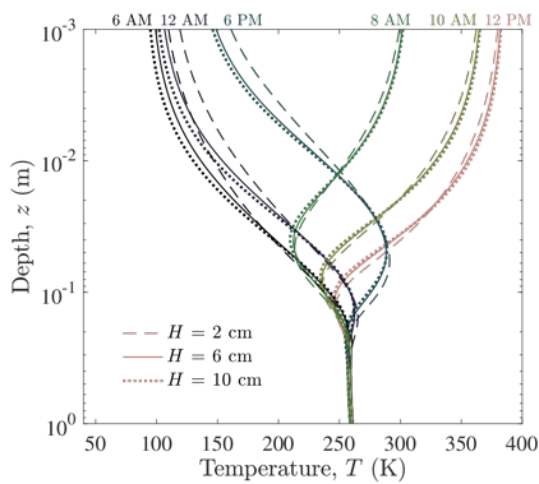

Figure 2: Depth profiles of model temperature for three different values of the regolith *H*-parameter: 2 cm (dashed curve), 6 cm (solid), 10 cm (dotted). Several different local times are shown: 12 AM, 6 AM, 8 AM, 10 AM, 12 PM, 6 PM.

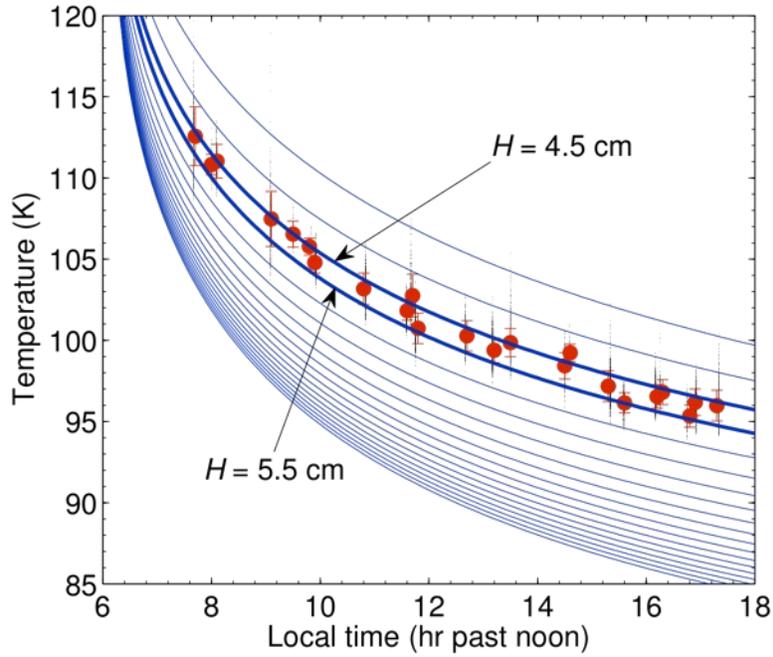

Figure 3: Model nighttime surface temperature curves for a 1×1 km location near Kepler crater (8.5°N, -38.5°E). Large points with error bars indicate mean and 1-σ deviations in the Diviner regolith temperature data (small black points) acquired on individual orbits. Solid curves are model temperatures for different values of the *H*-parameter, incremented by 1 cm. In this example, a best fit value of ~5 cm is obtained.

where $\rho_s$ is the density at the surface, and $\rho_d$ is the density at depths $z \gg H$. The H-parameter governs the increase of density and conductivity with depth. Figure 1 shows some example depth profiles. Very little experimental data exist to constrain the variation in thermal conductivity of granular materials in vacuum. Measurements of particulate basalt by Fountain and West (1970) suggest that over the relevant regime of density and temperature, thermal conductivity increases approximately linearly with density. For simplicity, we therefore prescribe conductivity as the dependent variable, as detailed in the Appendix. In this formulation, the parameter $H$ is treated as the only independent variable, and affects the overall profiles of both density and conductivity; see Section 3.3.2 for a discussion of alternative models. Since typically $\rho_d > \rho_s$, larger values of the H-parameter correspond to more insulating material near the surface, and smaller values correspond to denser, more conductive material. These properties affect the penetration of the diurnal thermal wave into the subsurface (Fig. 2). We performed model fits on the Diviner nighttime regolith temperature data with $H$ as the free parameter, given fixed values of $\rho_s$ (1100 kg m$^{-3}$) and $\rho_d$ (1800 kg m$^{-3}$), and the thermal conductivity of the surface and deep layers (7.4×10$^{-4}$ and 3.4×10$^{-3}$ W m$^{-1}$ K$^{-1}$, respectively). These parameter values are discussed in detail in the Appendix. Figure 3 shows an example set of model curves for one location with typical Diviner coverage on multiple orbits, over a region of interest ~1×1 km in size.

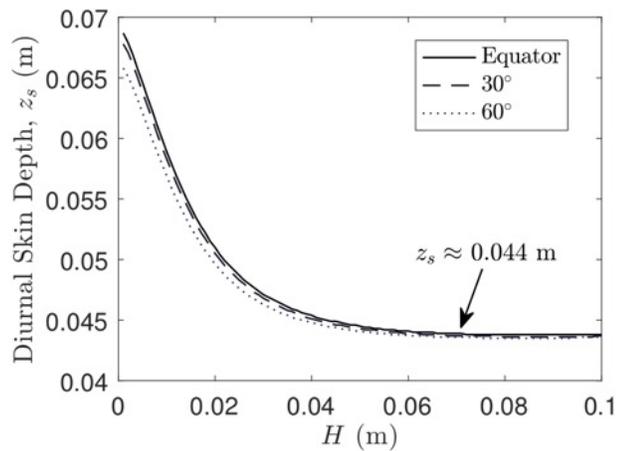

Figure 4: Variation of diurnal skin depth with H-parameter for three different latitudes, calculated from the numerical model, as the depth to 1/*e* damping of the surface temperature oscillations.

Thermal inertia is a quantity describing the resistance of materials to changes in temperature: $I = \sqrt{K\rho c_p}$. It has been used widely in planetary science, because it is related to the intrinsic physical properties of surface and subsurface materials affecting remote measurements of surface temperature variations (e.g., Mellon et al., 2000; Putzig and Mellon, 2007; Bandfield et al., 2008). In our present formulation, thermal conductivity varies with depth and temperature. Heat capacity depends upon the mass of material in each layer, and also the temperature (Ledlow et al., 1992). Thermal inertia is therefore a depth- and time-dependent quantity: density varies with depth, heat capacity varies with time (i.e., temperature), and conductivity varies with both depth and time. Surface temperature cycles used to infer thermal inertia are controlled by $K(\rho,T)$, $\rho(z)$, and $c_p(T)$ of the thermally active layer, $z_s$. To simplify this picture and capture the bulk properties of the upper regolith layer, we define a depth-averaged thermal inertia,

$$I_s(t) = \frac{1}{z_s} \int_0^{z_s} I(z,t) dz \qquad (2)$$

where the skin depth $z_s$ is pre-computed as the depth where the amplitude of the diurnal temperature oscillations decay to $1/e$ of their value at the surface. Defined in this way, the skin depth is only weakly sensitive to latitude, but varies from ~4.4 cm for large values of $H$, and up to ~7 cm for $H = 0$ (Figure 4). Figure 5 shows the relationship between the H-parameter and $I_s(t)$, at local noon, midnight, and a diurnal average calculated over the lunar rotation period $P$:

$$\langle I_s \rangle = \frac{1}{P} \int_0^P I_s(t) dt \qquad (3)$$

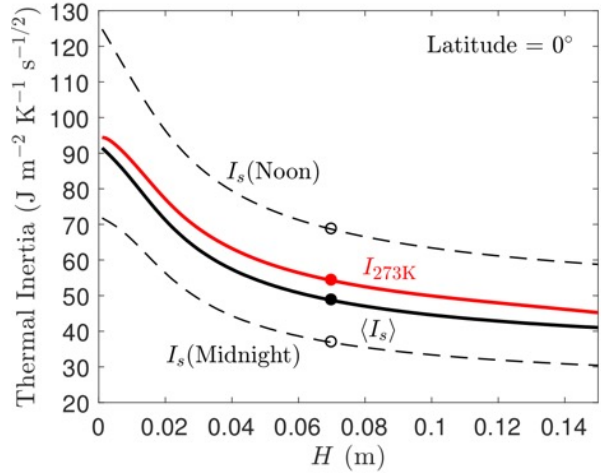

Figure 5: Dependence of regolith thermal inertia on the H-parameter, for a location on the equator. Although the curve for $\langle I_s \rangle$ varies with latitude and albedo due to the intrinsic temperature dependence of $K$ and $\rho$, the isothermal $I_{273K}$ curve does not. Circles indicate approximate average values for the whole Moon derived in this study. These models use thermal conductivity and density values: $K_s = 7.4\times10^{-4}$ W m$^{-1}$ K$^{-1}$, $K_d = 4.3\times10^{-3}$ W m$^{-1}$ K$^{-1}$, $\rho_s = 1100$ kg m$^{-3}$, $\rho_d = 1800$ kg m$^{-3}$, albedo $A_0$ = 0.12, and other standard parameters in Appendix A.

Given the temperature dependence intrinsic to $I_s(t)$, there is not a one-to-one correspondence between $H$ and $\langle I_s \rangle$. Instead, $\langle I_s \rangle$ depends on $H$, latitude, and albedo (through its effect on surface temperature). To remove the latitude- and albedo-dependence, we lastly define a thermal inertia at fixed temperature, $T_0 = 273$ K, and albedo $= 0.12$:

$$I_{273K} = \frac{1}{z_s} \int_0^{z_s} I(z, T_0) dz = \frac{1}{z_s} \int_0^{z_s} [K(T_0) \rho c_p(T_0)]^{\frac{1}{2}} dz \qquad (4)$$

With this definition, there is a one-to-one relationship between $I_{273K}$ and $H$. Furthermore, if thermal conductivity is assumed to increase approximately linearly with density (e.g., Fountain and West, 1970), then $I_{273K} \sim (K\rho)^{1/2} \sim \rho$. In other words, thermal inertia at a fixed temperature is roughly proportional to the bulk regolith density in the upper few centimeters.

### 3.2 Mapping

To generate the maps that follow, models were fit to the data falling within each 1/128-degree spatial bin. The number of data points within each bin was typically ~5 – 10, roughly randomly distributed over the local times 19:30 – 05:30 used. Lunar local times within 1.5 hr of sunset were excluded due to the persistent and complex effects of topography. Global maps of the H-parameter (and corresponding thermal inertia at 273 K) extending from the equator to ±70° were generated from fits to the regolith temperature data. Above ~60° latitude, artifacts introduced by topography become problematic. Several regions of interest were also investigated based on their geologic setting or atypical properties seen in the global context. However, many important or unusual features have not yet been studied in detail.

*3.2.1 Slope correction*

Daytime surface temperatures on the Moon are strongly influenced by topographic slopes, which cause small but noticeable perturbations to nighttime regolith temperatures. We applied a correction to these temperatures based on local slope $x$ (dimensionless) and azimuth angle $\gamma$ (radians from north) computed from the 128-ppd gridded LROC digital elevation model. In the temperature calculations, local times $t$ (lunar hr = 0 to 24) were adjusted based on the east-west component of the slope,

$$t' = t + \frac{12 \text{ hr}}{\pi} \tan^{-1}(x \sin \gamma) \qquad (5)$$

and latitudes $\phi$ were adjusted based on the north-south component of the slope,

$$\phi' = \phi + \frac{180°}{\pi} \tan^{-1}(x \cos \gamma) \qquad (6)$$

Care must be taken to check that the resulting local times and latitudes fall within the valid ranges of 0 to 24 hr and -90° to +90°, respectively. The slope corrections produce accurate results up to ~60° latitude; shadowing and insolation patterns become more complex at higher latitudes.

*3.2.2 Albedo correction*

To account for the effects of surface albedo on nighttime temperatures, we first produced a global bolometric Bond albedo map. This was done using: 1) the gridded 10-ppd LOLA 1064-nm normal albedo (Lucey et al., 2014), and 2) Diviner bolometric solar channel measurements (Vasavada et al., 2012). Although the LOLA data were acquired in a narrow 1064-nm wavelength band, they have the advantage of being acquired at zero-phase angle over the whole globe, largely eliminating artifacts due to topographic shadowing. The Diviner solar channel data provide the more accurate measurement of broadband solar albedo, which is the critical quantity in thermal calculations. To generate the global map, we scaled the LOLA data to match the Diviner equatorial reflectance, where topographic effects are minimized. The scaling factor was $f = 0.49$, converting from the LOLA normal albedo, $A_L$, to the equivalent broadband solar albedo, $A_\odot = f A_L$. Using this approach, we found point-to-point RMS differences between the Diviner equatorial albedo and $A_\odot$ of < 0.01. Our resulting bolometric albedo model is shown in Section 4.1. Although important, albedo variations on the lunar surface result in a modest effect on the derived thermophysical properties, as demonstrated in Section 3.3.3.

**3.3    Uncertainties and caveats**

Several considerations should be made in order to interpret the results below. First is the theoretical distinction between "regolith" and "rock". Second, our model necessarily contains assumptions that must be examined closely in order to estimate the errors and explore plausible alternative models. Third, we must account for the propagation of measurement errors and uncertainties introduced by the steps in processing and fitting the data. Here we discuss each of these points.

*3.3.1    Rock vs. regolith*

The term *regolith* is defined as a "layer or mantle of fragmental and unconsolidated rock material" (McKay et al., 1991). Any distinction between rocks and regolith is therefore arbitrary, so long as the "rocks" are discontinuous with underlying bedrock. Nonetheless, we will show below that the regolith temperatures derived from Diviner's multi-spectral measurements reveal patterns distinct from the distribution of thermally detectable rocks on the lunar surface. In part, this difference is explained by the relatively small (< 1%) population of larger rocks > 10 – 100 cm globally (Bandfield et al., 2011), and the pervasiveness of regolith. It should be understood that the regolith thermophysical properties derived here include the contributions of rock fragments smaller than half a meter.

*3.3.2    Model assumptions and alternatives*

For simplicity and consistency with earlier models and *in situ* measurements from the Apollo missions, we have assumed the exponential profile of equation (1) applies everywhere on the Moon. Further, the values of $\rho_s$ and $\rho_d$, $K_s$ and $K_d$, are assumed fixed, with the vertical variation in thermophysical properties specified by the parameter $H$. It must be acknowledged that variations in all of these parameters occur, and therefore the present model is a simplification. However, thermal inertia is the fundamental property controlling the nighttime regolith temperatures, so variations in thermal inertia arising from different thermophysical properties ($K$, $\rho$, $c_p$) will manifest as variations in best-fit H-parameter values. Therefore, this model provides a quantitative estimate of the thermal inertia over the upper few thermal skin depths (~10 cm), which reflects the bulk thermophysical properties of interest for a range of studies. As demonstrated below, the model is useful for quantifying and interpreting these upper regolith properties.

Alternative models may be formulated to fit the Diviner data. For example, keeping equation (1), it may be possible to fit more than one free parameter, given a sufficient number of measurements at each location. We attempted multivariate fitting using non-linear least-squares minimization to simultaneously fit $K_s$ (or $\rho_s$) and $H$ for several test regions. However, we found that these models were ill-conditioned; small changes in the surface boundary condition ($K_s$ and $\rho_s$) result in large changes in the best-fit $H$-parameter, with little effect on the residuals. In other words, the fits were non-unique. Therefore, we chose to fix the upper boundary condition to values consistent with previous work (see Appendix for further discussion). Future work using more data, especially early in the night, may be more successful performing a multi-variate fit to map other parameters.

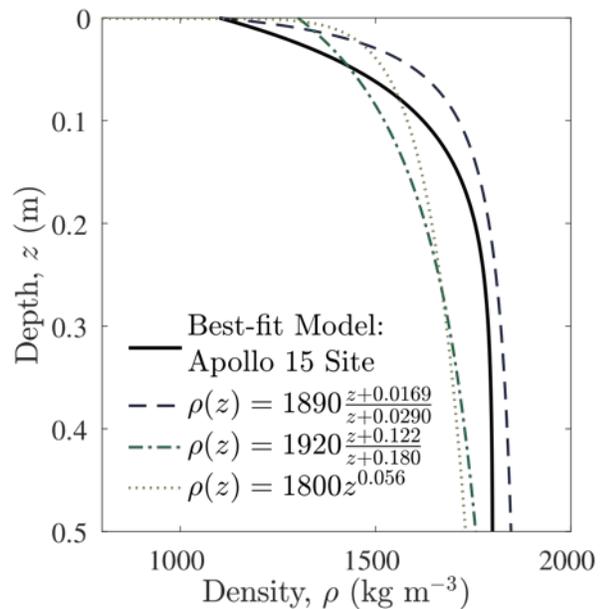

Figure 6: Comparison of vertical density profile derived in this study for the Apollo 15 landing site (solid curve), with those given (dashed and dotted curves) by *Carrier et al.* (1991). The constraints used by *Carrier et al.* to derive the latter fits apply over discrete depth intervals, and are given in Table 1.

*3.3.3  Validation, uncertainties and error estimation*

Density profiles determined from Apollo core tubes and *in situ* analysis (Mitchell et al., 1973; Carrier et al., 1991) provide "ground truth" for our model (Fig. 6). Given the coarse vertical resolution of the in-situ and core tube measurements, our density profiles are in good agreement over all depths sampled. Table 1 summarizes the comparison of our best-fit density profiles to the Apollo-era data. Carrier et al. (1991) also presented three different fits to these data, including one power-law and two hyperbolic fits. However, we found no significant difference in the quality of these fits, given the coarse spatial sampling, and therefore include all three of them for comparison in Fig. 6.

Various sources of temperature differences on the lunar surface may contribute to errors in the model results presented here, including sub-pixel spatial variations in thermophysical properties, vertical layering, surface roughness, and re-radiation by warm rocks. Sub-pixel variations can be addressed by interpreting the derived thermal inertia values as a spatial average over the given grid scale (typically 1/128-degree ~ 250 m), although it should be understood that this average is weighted towards higher-inertia materials, due to their higher nighttime temperatures. Rock-regolith heating is an important consideration for surfaces with rock abundance >3%, as shown in Appendix B, and discussed in more detail in Section 4.1.2.

Formal errors on regolith thermal inertia due to Diviner's measurement uncertainty are small. Over the range of nighttime temperatures used in this study, ~60 – 120 K, Diviner has a noise-equivalent temperature difference (NETD) $\delta T$ < 1 K (Paige et al., 2010a). This is a conservative estimate, because at 120 K, the NETD is ≪ 1 K, and binning a number of measurements $N$, reduces the error by $\sim N^{-1/2}$. Random measurement errors in surface temperature lead to expected uncertainties in thermal inertia with magnitude

$$\delta I = \frac{\partial I}{\partial H} \delta H = \frac{\partial I}{\partial H} \frac{\partial H}{\partial T} \delta T \qquad (7)$$

Thermal model results over a range of $H$-parameter values indicate $\partial H / \partial T \approx 0.5 - 1$ cm K$^{-1}$ and thermal inertia $\partial I / \partial H \approx 2 - 4$ J m$^{-2}$ K$^{-1}$ s$^{-1/2}$ cm$^{-1}$ at midnight local time, yielding a relative error of $\delta I / I \approx 4 - 6\%$ for $\delta T = 1$ K. Other sources of temperature error, such as slopes and surface roughness, can be propagated using the formula above and the given derivatives.

Uncertainties in the bolometric Bond albedo of the lunar surface could lead to errors in the derived thermophysical properties. We estimated the magnitude of these errors by calculating model surface temperatures (at local midnight) over a range of albedo values. The temperature error introduced by an albedo difference $\delta A$ is $\delta T = (\partial T / \partial A) \delta A$, and we find $\partial T / \partial A \approx 0.2$ K/%, where albedo is given in percent reflectance. The lunar surface albedo varies from ~5% in the maria to ~25% in the highlands (Section 3.2.2; Vasavada et al., 2012).

Our knowledge of this variation based on global measurements (e.g., Lucey et al., 2014) constrains the uncertainty in the bolometric albedo to < 5%, such that we expect temperature uncertainties of < 1 K, and relative thermal inertia errors of < 6%.

Table 1: Constraints on lunar regolith density profiles. Error bars on the values derived in this study correspond to the 1-σ range in fits to the Diviner data.

|  | Density (kg m$^{-3}$) | |
| --- | --- | --- |
| Depth Range (m) | Apollo *in situ*[a] | This study |
| 0 – 0.15 | 1500 ±50 | 1530 ±100 |
| 0 – 0.30 | 1580 ±50 | 1630 ±50 |
| 0.30 – 0.60 | 1740 ±50 | 1790 ±10 |
| 0 – 0.60 | 1660 ±50 | 1710 ±50 |

[a]Mitchell et al. (1973)

## 4 Results

Regolith properties derived from the Diviner data exhibit global and regional patterns, with expressions of important geophysical, geological, and geochemical processes, such as: impacts, volcanism, and interaction with the space environment. Many of these features can be readily understood based on earlier work, while others require further study. Below we present some of the general patterns in the global maps (Fig. 7 & 8), and then briefly describe some of the more prominent features at the regional and local scales.

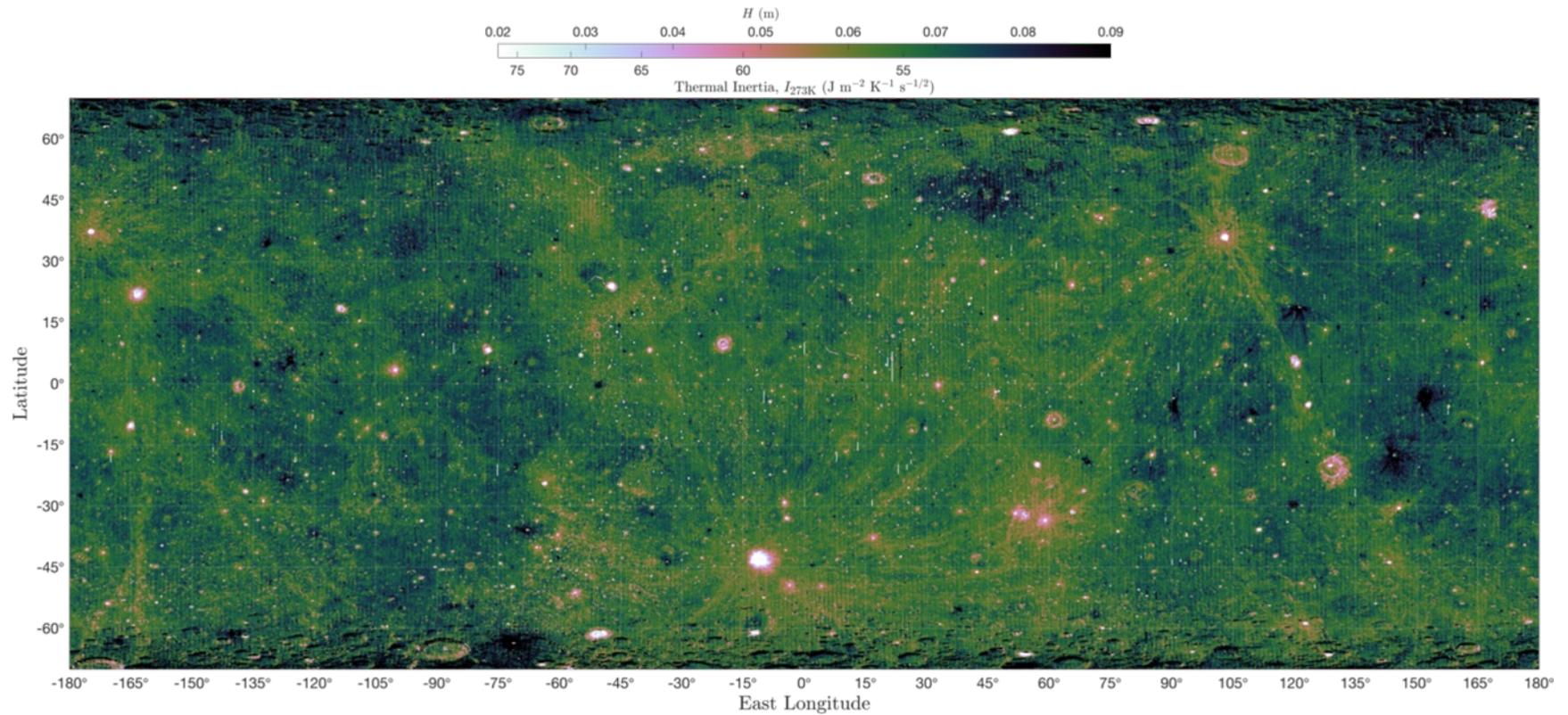

Figure 7: Global map of the H-parameter and equivalent regolith thermal inertia $I_{273K}$. The map shown spans ±70° latitude, ±180 longitude, with data averaged in 1/10-degree spatial bins. The color scale also indicates the regolith *H*-parameter, in meters. Bright features are typically fresh impact craters < 1 Ga, many showing high-inertia proximal ejecta and extensive rays. Dark features typically fall into two categories: "cold spots" (very recent impact features described by *Bandfield et al*., 2014), and regional pyroclastic deposits. However, many features have an ambiguous origin; for example, the extensive low-thermal inertia feature surrounding Atlas crater near +45°N, +45°E.

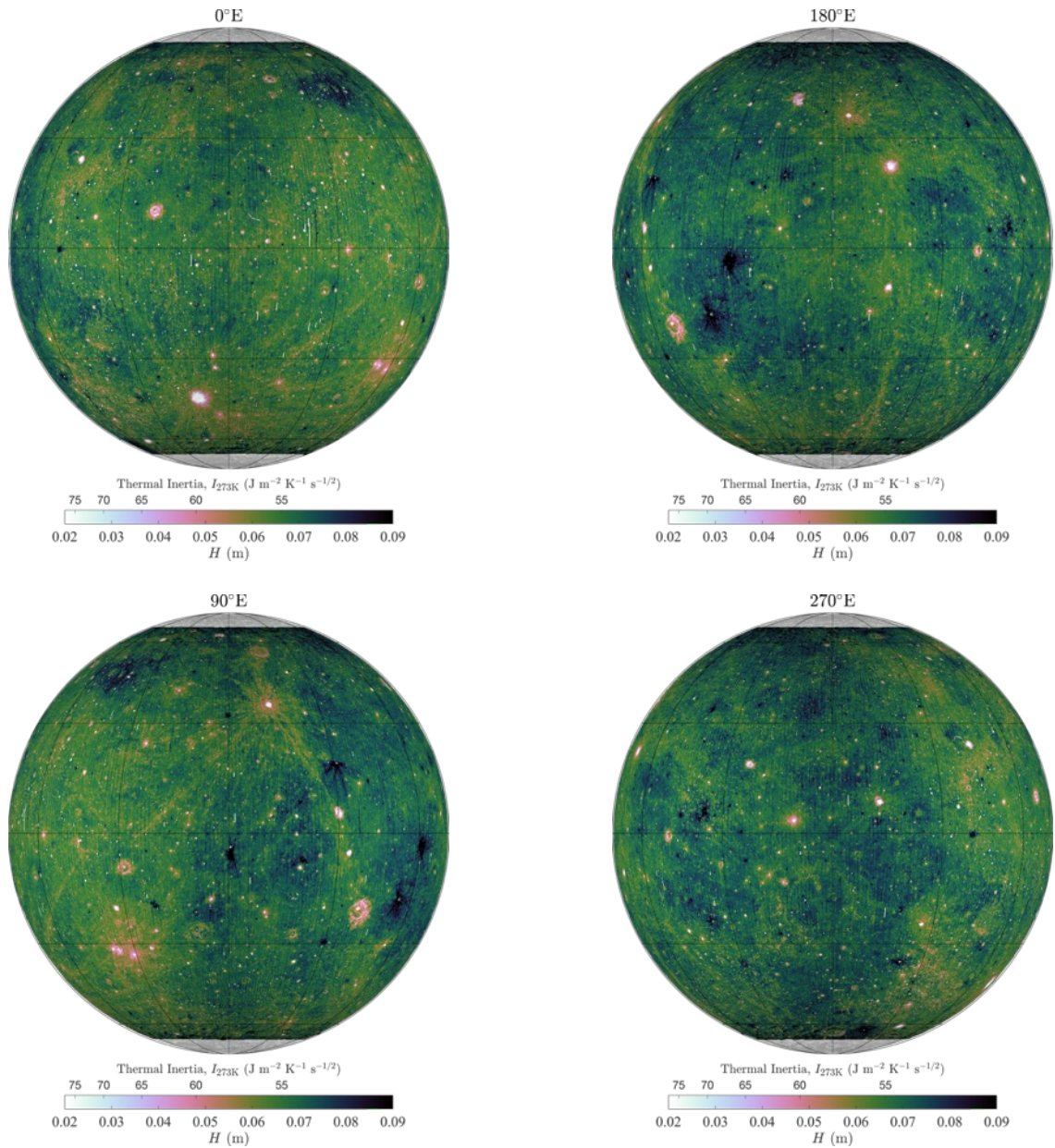

Figure 8: Orthographic projections of the H-parameter and equivalent regolith thermal inertia $I_{273K}$ for the nearside (0°E, upper left), farside (180°E, upper right), trailing hemisphere (90°E, lower left), and leading hemisphere (270°E, lower right). Latitude and longitude lines are spaced at 30°. Dark splotches are typically "cold spots" surrounding very young impact craters, whereas the brightest features are large impact craters and their ejecta. Many, though not all, of the more prominent pyroclastic deposits have low thermal inertia. Some of the largest crater rays are seen to wrap around the globe (notably those of Tycho crater at ~43°S, -11°E). The prominent rayed crater in the 90°E projection is Giordano Bruno.

## 4.1  Global

On the global scale, the lunar regolith exhibits remarkable uniformity, without a prominent maria/highlands or nearside/far-side dichotomy. Patterns at this scale are dominated by higher thermal inertia ejecta from recent impact craters > 10 km in size (e.g., Tycho, 43°S, -11°E), contrasting with lower thermal inertia in regions with fewer recent large craters. Although the nearside maria do not stand out distinctly, several of the maria (e.g., Humorum, Procellarum, and Frigoris) exhibit a concentration of higher-thermal inertia materials surrounding smaller craters. This may be due to the younger surface age and hence thinner regolith layer.

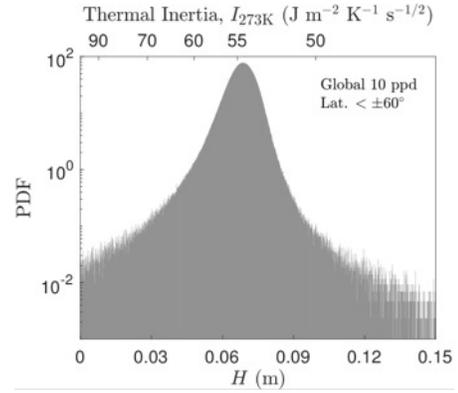

Figure 9: Global probability distribution function for regolith H-parameter and equivalent thermal inertia, binned spatially at 10 pixels per degree. The mean values are 0.068 m, and 55 J m$^{-2}$ K$^{-1}$ s$^{-1/2}$.

Prominent higher thermal inertia features >30 km in size also appear in the maps of Yu and Fa (2016), who independently fit the surface conductivity term, $K_s$. We suggest that the strong latitude gradient in their $K_s$ map may be due to solar incidence-dependent albedo effects, which complicate independent retrieval of conductivity (see Appendix A). After removing this gradient, Yu and Fa (2016) noted generally higher thermal conductivity material in the highlands (~1.8× higher $K_s$), which they attributed to smaller regolith grain sizes. The fact that this dichotomy does not appear in our regolith H-parameter maps is an intriguing result, which we discuss in Section 5.

At smaller spatial scales, many striking features appear. For instance, distinct, extended low-thermal inertia patterns typically surround very fresh impact craters < 10 km in diameter. These ubiquitous features are called "cold spots" (Bandfield et al., 2014), and are discussed in detail below.

In general, the regolith model fits the observations well over the whole lunar surface, with root-mean-square (RMS) deviations of the best-fit model temperatures of < 1 K from the measurements. Derived values of the H-parameter have global (latitude ±60°) average and mode values of 6.8 cm, with a standard deviation of ~0.7 cm. These values correspond to a globally averaged thermal inertia $I_{273K} \approx 55$ with standard deviation ~2 J m$^{-2}$ K$^{-1}$ s$^{-1/2}$ (Figure 9). This is similar to thermal inertia values for the bright unconsolidated fines units in the equatorial regions of Mars (Putzig et al., 2005). Regoliths on Main-Belt and Near-Earth asteroids typically range from ~10 to 100 J m$^{-2}$ K$^{-1}$ s$^{-1/2}$ for bodies with diameters larger than ~10 km (Delbo and Tanga, 2009), with estimates for Ceres (Spencer, 1990) and Vesta (Capria et al., 2014) roughly 10 and 30 J m$^{-2}$ K$^{-1}$ s$^{-1/2}$, respectively. However, some caveats are necessary when comparing lunar thermal inertia values to those of other airless bodies. With their much shorter rotation periods, diurnal skin depths on asteroids are ~10x shallower (< 1 cm) than on the Moon. Therefore, thermal emission measurements of asteroids typically probe depths where the lunar thermal inertia would be ~2x lower than the bulk average. Asteroid thermal models typically fit day-side temperatures, and derived thermal inertia values represent the bulk material, including both rocks and regolith. Delbo et al. (2015) discuss these complexities in detail, and present some possible temperature corrections at different heliocentric distances to account for known effects.

*4.1.1    Comparison to Albedo and Optical Maturity*

Lunar regolith becomes darker and redder as it "matures" with exposure to the space environment (Fischer and Pieters, 1996). Lucey et al. (2000) derived a quantitative optical maturity index (OMAT) based on observed spectral changes with age, which can be used to infer relative exposure ages of lunar surface materials. Although OMAT is known to be influenced by mineral composition and geologic setting, thermal inertia is not directly affected by composition (Fountain and West, 1970; Wechsler et al., 1972). Therefore, it may be possible to separate maturity from intrinsic composition using thermal inertia and OMAT together.

Visual comparisons to OMAT (Fig. 10) show that many of the prominent optically immature (high OMAT) craters and their ejecta also appear as higher-thermal inertia deposits in the Diviner maps. Conversely, some of the older surfaces, for example the "South Pole-Aitken Basin" (~53°S, -169°E), exhibit lower thermal inertia overall. This can be understood as the natural breakdown of rocks and regolith grains by the continuous meteoritic flux at the lunar surface. When larger impacts penetrate through the regolith layer, they may excavate bedrock (or buried rocks) and raise the thermal inertia of the surrounding regolith. Over time, these rocky regions are pummeled and pulverized, lowering the thermal inertia. However, the quantitative correlation between OMAT and thermal inertia is weak (Fig. 11), with an estimated Pearson correlation coefficient $R \approx 0.07$. The correlation with visible albedo is even lower. Again, this behavior is consistent with laboratory work showing that composition plays a minor, indirect role in determining the thermophysical properties of granular materials (Wechsler et al., 1972).

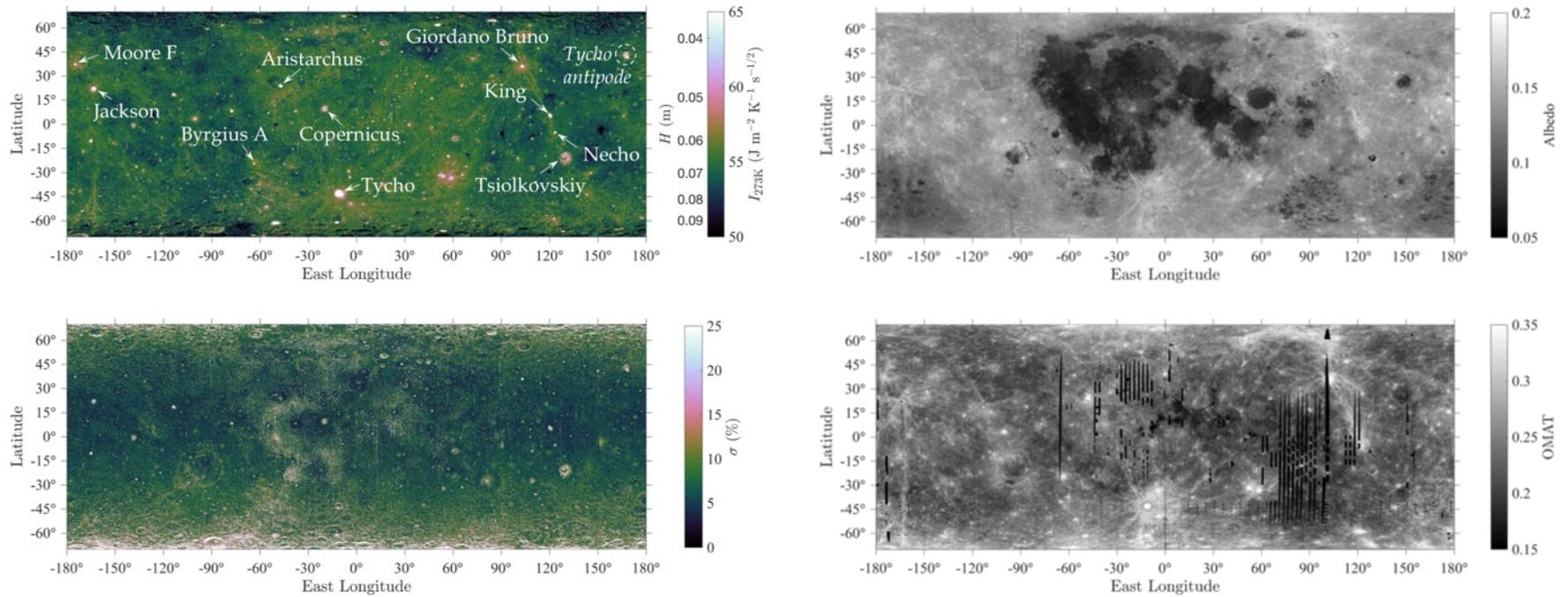

Figure 10: Regolith thermal inertia $I_{273K}$ (upper left panel) and its spatial variance at 10 ppd (lower left panel), compared to LOLA normal albedo (upper right panel) and optical maturity (lower right panel). Note that in contrast to Fig. 7-8, the color scale in the upper left panel is linear for $I_{273K}$ and is therefore logarithmic for the equivalent H-parameter. By convention, the optical maturity parameter (OMAT) is higher for less mature surfaces.

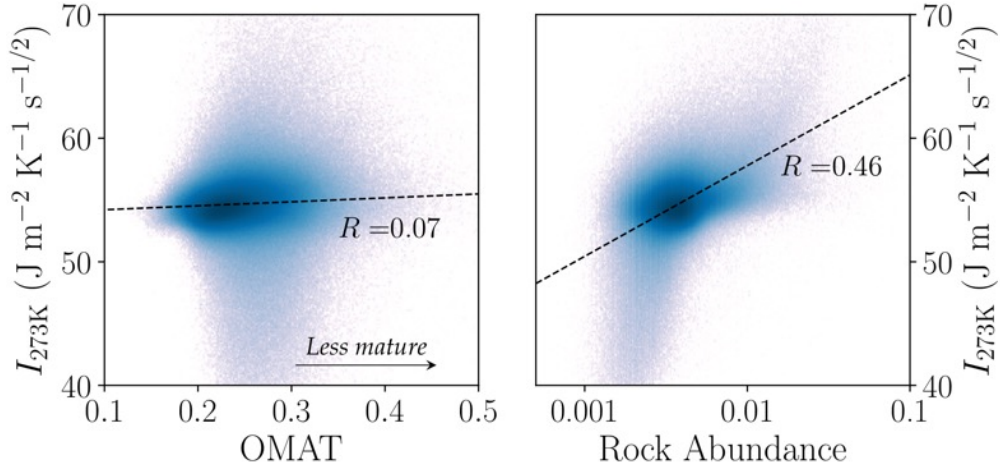

Figure 11: Cross-plots of thermal inertia, $I_{273K}$, versus optical maturity (OMAT, left) and rock abundance (right), generated from global +/-70° latitude maps at 10 ppd resolution. Shading indicates the number of data points within each bin, with a logarithmic scale from 1 (light gray) to $10^4$ (dark blue). Thermal inertia is positively correlated ($R$ = 0.46) with rock abundance, which can be explained by the presence of small rocks on the surface and mixed into the regolith, and/or radiant heating by larger rocks. Optical maturity shows only a weak correlation ($R$ = 0.07) with thermal inertia, where higher OMAT values (less mature) correlate with higher thermal inertia. This behavior is consistent with fresh impact craters, which expose rocky, immature materials, which may be quickly weathered in the space environment.

*4.1.2 Comparison to Rock Abundance*

Rock abundance refers to the areal concentration of rocks larger than ~10 – 100 cm derived from Diviner's nighttime multi-spectral infrared measurements (Bandfield et al., 2011). For the present study, we used rock abundance and regolith temperatures derived from available gridded Diviner data, using the same dataset as in Bandfield et al. (2017). As described earlier, the regolith temperatures used to derive H-parameter are the result of the same initial fitting procedure used for rock abundance. In effect, the regolith temperatures are the residual thermal emission left after removing the contribution of the rocks (though the rocks may be minor contributors). Our model then fits this thermal emission curve by varying the H-parameter to minimize the second-order residuals and find the best-fit profile. Any buried rocks would not affect the Diviner rock abundance measurement, but would increase the regolith thermal inertia if they were < 10 cm from the surface (Elder et al., 2017a).

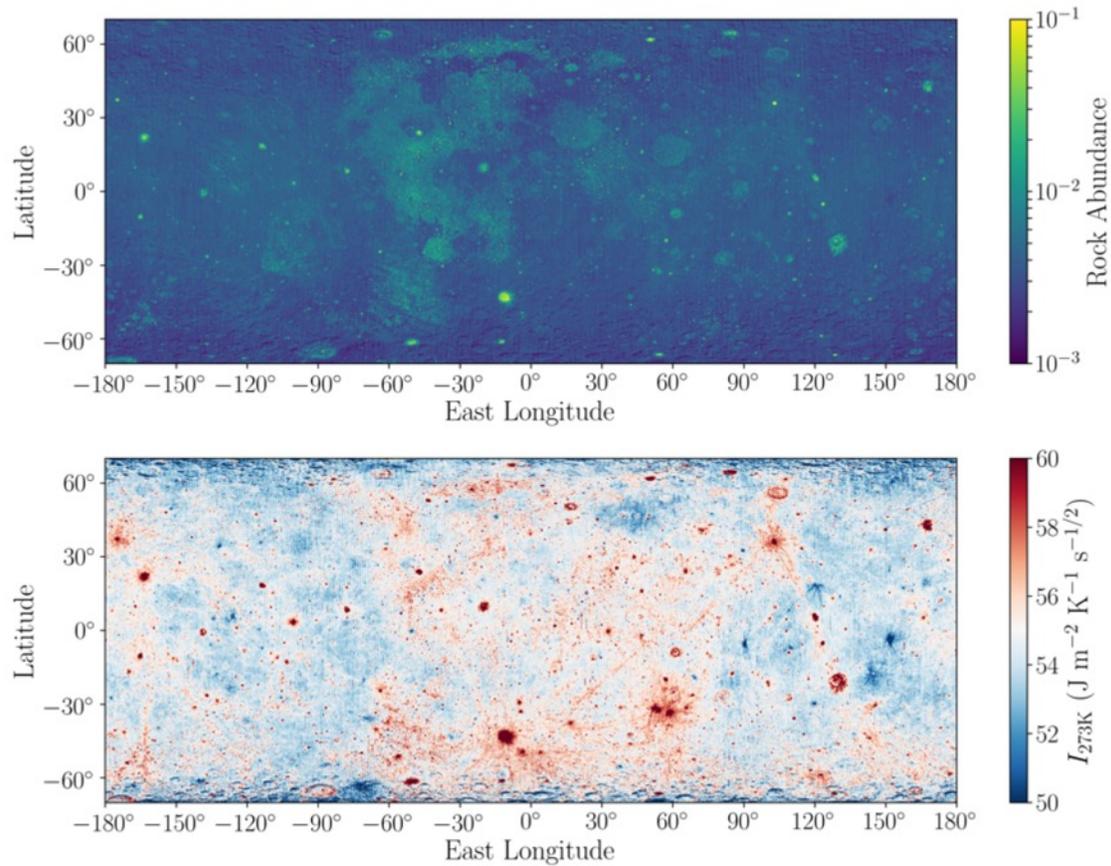

Figure 12: Rock abundance (upper panel) and thermal inertia $I_{273K}$ (lower panel), both binned at 10 ppd. The color scale shown for $I_{273K}$ is intended to highlight both high- and low-thermal inertia features, which appear dark red and dark blue, respectively.

Regolith H-parameter and thermal inertia might be expected to correlate with rock abundance for at least two reasons: 1) small rocks (not included in the rock abundance measure) are often present in high concentrations where larger rocks occur, and 2) conduction and re-radiation from rocks may augment nighttime regolith temperatures in rocky regions. Our results show that rock abundance is only correlated with thermal inertia in some of the rockiest regions (Fig. 11 and 12). This is expected based on models (Appendix B), which show that rocks heat surrounding regolith by infrared radiation and conduction, but the effect is only detectable for rock abundance >3%. It may become dominant for rock abundance >10%, in which case the effect must be explicitly modeled. Only a small fraction of the lunar surface exhibits concentrations >3% of meter-scale rocks, where we indeed see a stronger correlation between regolith thermal inertia and rock abundance (Fig 11). Further work is needed to carry the analysis further, because unmapped small rock fragments are more prevalent than meter-scale rocks, and may also contribute to rock-regolith heating. Here, we simply note that rock-regolith mutual heating involving rocks >1 m does not affect our derived regolith H-parameter for a large majority of the lunar surface.

Many of the more prominent features in the regolith thermal inertia maps do not show a prominent rock abundance signature. For example, in the rock abundance maps, the nearside maria glow with the presence of large surface rocks, yet their boundaries appear indistinct in the regolith thermal inertia maps. Unique patterns in the thermal inertia maps also emerge, including cold spots and very prominent crater rays. Some of these features are described briefly in the following sections.

**4.2  Impact Craters and Ejecta**

Previous work indicated that rock abundance is a key indicator of crater age (Bandfield et al., 2011; Ghent et al., 2014). Ghent et al. (2014) demonstrated that a quantitative relationship exists between the 95$^{th}$ percentile rock abundance within a crater's ejecta blanket, and its model age based on counts of superposed craters. This result showed that rock breakdown at the lunar surface proceeds at an unexpectedly rapid rate initially, followed by a tapering off as rocks are removed by comminution. Here, we investigated whether a similar relationship exists between model crater ages and thermal inertia, which is expected since rock breakdown is coupled to regolith formation.

Figure 13 displays regolith thermal inertia maps of four prominent Copernican-aged craters: Giordano Bruno, Moore F, Aristarchus, and Copernicus. These are four of the nine craters considered by Ghent et al. (2014). A general time-evolution is apparent, from higher thermal inertia (lower H-parameter) in the ejecta and interiors of younger craters, to lower thermal inertia (higher H-parameter) approaching that of the background, for older craters. To quantify this trend, we measured regolith thermal inertia at various radial distances from each crater's rim (Fig. 14). The results show that the greatest changes occur within ~2 crater radii during this 1 Myr – 1 Gyr period after crater formation. A power-law model fits the data well:

$$\bar{H} = H_0 t_{\text{Ma}}{}^b \qquad (8)$$

where $\overline{H}$ is the average regolith H-parameter from the crater rim to 1.5 crater radii, with $H_0 = 0.032$ m, $b = 0.10$, and $t_{\text{Ma}}$ the crater age in millions of years. Given an observed H-parameter average for an ejecta blanket, this equation can be inverted to give an estimate of the crater's age. Increased robustness of this age-dating approach may be possible when combined with the technique presented by Ghent et al. (2014) using rock abundance. At least one crater bucks the age vs. H-parameter trend: Tsiolkovsky (20°S, 129°E) has a much lower H-parameter (higher thermal inertia) than its published >3.2 Ga age would indicate. Tsiolkovsky also has anomalously high rock abundance. These features are consistent with the interpretation of a massive impact-melt event coincident with its formation and/or subsequent modification by other impact processes (Greenhagen et al., 2016).

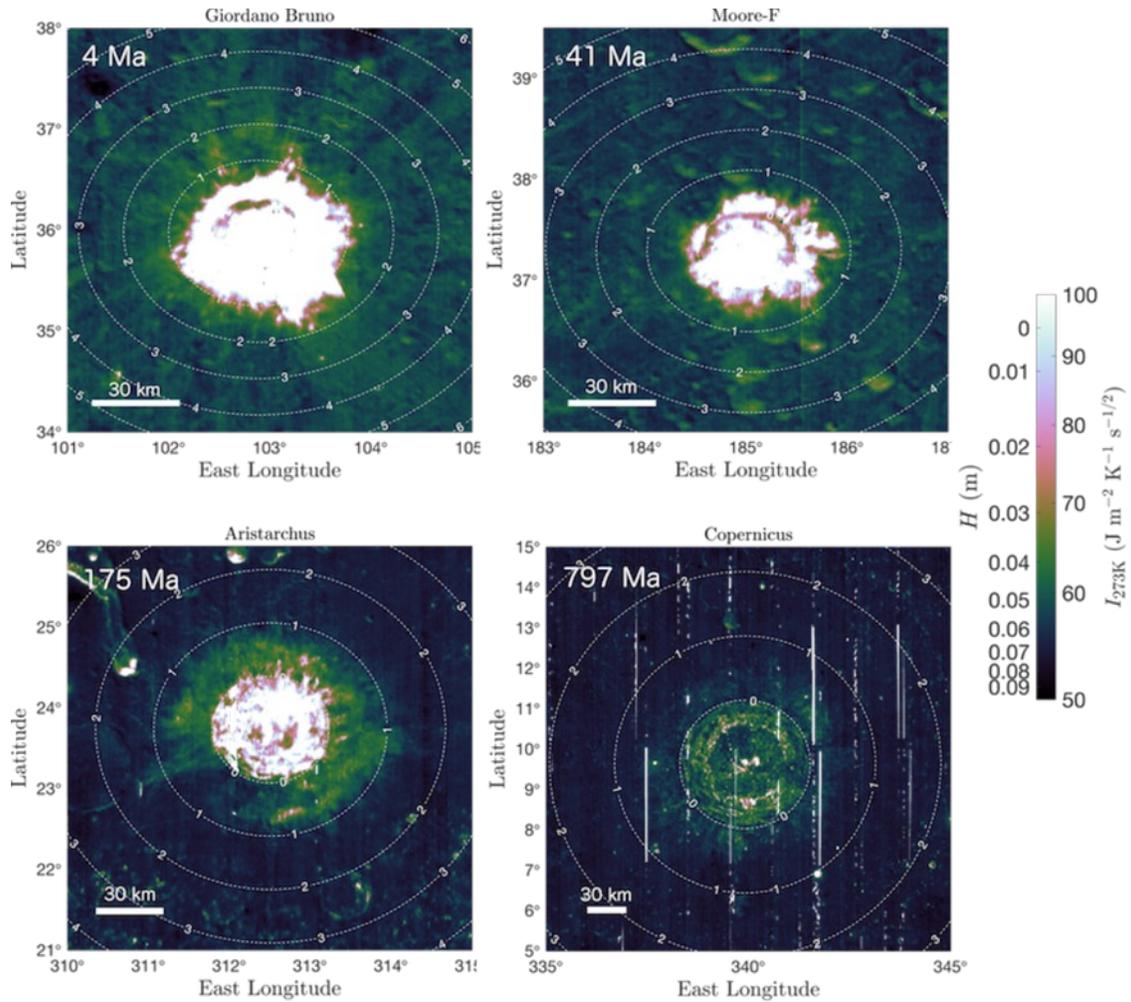

Figure 13: A progression of craters showing the decrease in thermal inertia (or increase in H-parameter) with age. The four craters shown are: Giordano Bruno (4 Ma), Moore F (41 Ma), Aristarchus (175 Ma), and Copernicus (797 Ma). Ages are published values based on crater statistics, and are taken from Ghent et al. (2014) and references therein. The dashed white lines show distances from each crater rim, marked in units of crater radii.

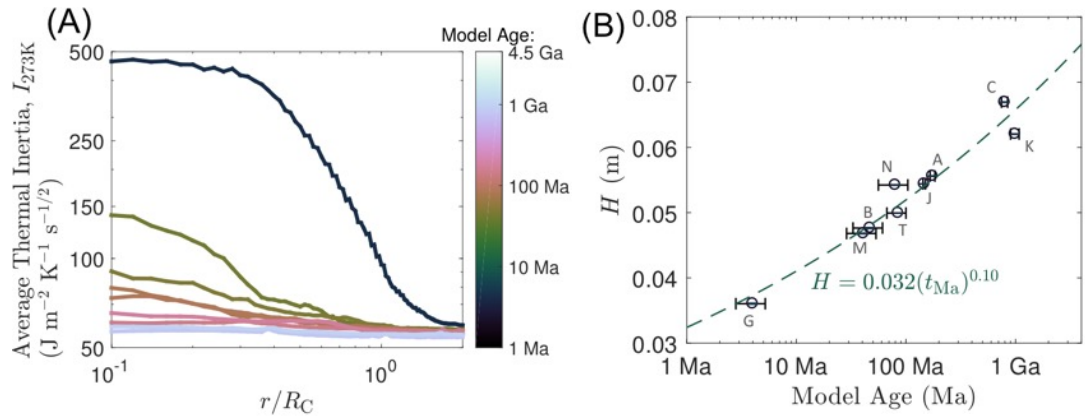

Figure 14: Regolith thermal inertia compared to crater age. (A) Each curve represents the average regolith thermal inertia $I_{273K}$ at varying distances $r$ from the crater rim, scaled to the crater radius, $R_C$. Colors indicate the model ages (see text) for each of the craters. (B) Average thermal inertia of crater ejecta from the rim to $2R_C$, for each of the nine craters investigated by Ghent et al. (2014): King, Copernicus, Aristarchus, Jackson, Tycho, Necho, Byrgius-A, Moore-F, Giordano Bruno. The curve in (B) is the result of a nonlinear least-squares fit to the data, using $H = H_0 t_{Ma}^b$, with best-fit parameters $H_0 = 0.032$ m, $b = 0.10$, where the crater age $t_{Ma}$ given in millions of years.

*4.2.1   Cold Spots*

Among the many intriguing features in the Diviner nighttime regolith temperature maps, the thousands of "cold spots" are perhaps the most striking. In the nighttime thermal maps, these prominent low-temperature, ray-like patterns surround very young craters, giving the impression of a splash of fluffy material. However, as noted by Bandfield et al. (2014), these features cannot be explained by the emplacement of ejecta alone; they extend to $10 - 100$ crater radii in many cases, where ejecta thicknesses should be negligible. Yet, their persistent thermal signature during the lunar night implies a modified layer at least ~5 cm thick. This thickness and their radial extent yield an estimated volume of material, which is typically much larger (>10x) than the crater volume. Therefore, the impacts forming their host craters have modified the regolith to much greater distances than expected. Closer to the crater rim, images from the LROC-NAC show flow-like morphologies consistent with energetic, granular flow, concentrated along the visible rays (Bandfield et al., 2014).

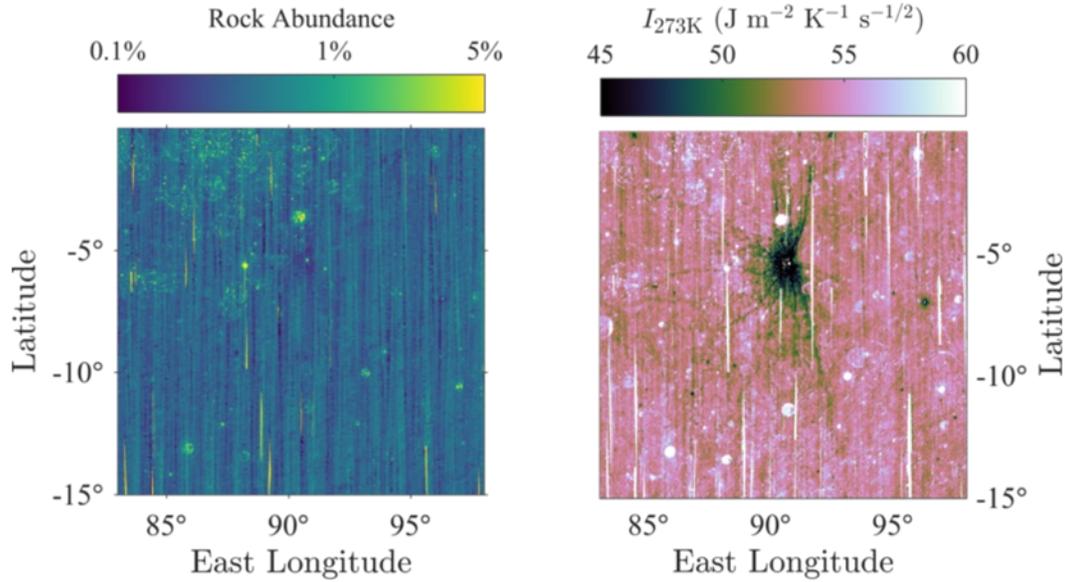

Figure 15: Rock abundance (left panel) and regolith thermal inertia (right panel) showing a large, ~300-km cold spot. Although the cold spot is much more prominent in the thermal inertia map, a very low rock abundance region is also apparent within ~10-20 km of the ~1 km diameter crater. The surface is very rocky within ~1 km of the crater rim. Several smaller cold spots are also seen in the thermal inertia frame, although it is not clear whether they display a similar pattern in rock abundance.

At the kilometer scale, cold spots typically exhibit $H > 0.1$ m, or thermal inertia values in the range $I_{273K} \sim 40 - 50$ J m$^{-2}$ K$^{-1}$ s$^{-1/2}$, which is lower than the global average regolith thermal inertia by ~10 – 30%. At smaller spatial scales, thermal inertia values can be even lower (< 20 J m$^{-2}$ K$^{-1}$ s$^{-1/2}$), especially in the largest cold spots. Since $I_{273K}$ is the mean value over the skin depth sampled by the diurnal thermal wave, this lower thermal inertia is understood to be a bulk property extending to several cm depths within the cold spot features. Eclipse data (Hayne et al., 2015b) indicate the uppermost ~1 – 10 mm of cold spots may not have distinct thermophysical properties; if anything, they may have higher thermal inertia in this uppermost layer. Paradoxically, the Diviner rock abundance derived from diurnal data (Fig. 15) appears to show a lack of rocks within ~10 crater radii of some of the larger cold spot craters. One possible resolution to this discrepancy, may be that small rocks < 10 cm (not measured by the standard Diviner rock abundance technique) are more prevalent in the cold spots, lying on top of a lower-density layer. A complete description of the nature and formation of cold spots remains elusive. Yet, the regolith thermal inertia maps presented here may be used to better constrain cold spot formation models, such as their size-frequency distribution (hence age), morphologies, radial decay with distance from the crater rim, and other important properties.

*4.2.2 Tycho Antipodal Deposit*

One of the more prominent thermophysical anomalies on the Moon is a rocky area located near the antipode of Tycho crater (Figure 16). Bandfield et al. (2017) suggested that this concentration of high-thermal inertia materials and melt deposits may be related to the formation of Tycho itself, which is of a similar inferred age to the antipodal deposit, ~20 – 100 Ma (Robinson et al., 2016). These authors identified directionality in the rocks and melt features, indicating the deposition of low-angle ejecta from the Tycho impact, concentrated at its antipode. The absence of large, Copernican-aged impact craters or other possible sources lends support to the Tycho antipode hypothesis.

The Tycho antipodal deposits also appear prominently in the regolith thermal inertia data (Fig. 16). Similar azimuthal asymmetries are noted as in the rock abundance, with isolated 1 – 10 km-scale contiguous regions with ~30 – 40% higher thermal inertia than the background. Minor differences between thermal inertia and surface rock abundance may be related to the presence of smaller rocks at the surface and in the subsurface. For example, we also note ray-like patterns extending to the NW, E, S-SE, and a possible faint feature extending to the NE, present in the thermal inertia map, but not the rock abundance map.

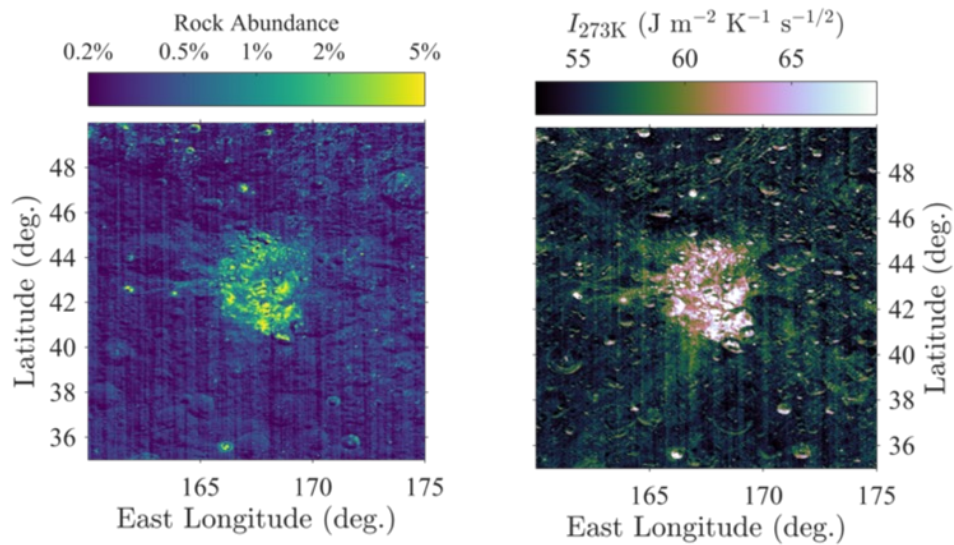

Figure 16: Rock abundance (left panel) and regolith thermal inerta (right panel) for the so-called Tycho antipode deposit (Robinson et al., 2016; Bandfield et al., 2017). We note similar patterns of rocky and higher-thermal inertia materials, although ray-like features are more prominent in the thermal inertia map.

**4.3 Pyroclastic Deposits**

Several regional pyroclastic deposits (>2500 km$^2$; Gaddis et al., 2000; Allen et al., 2012) are notable in the Diviner thermal inertia dataset. One of the largest is on the Aristarchus plateau, where both rock abundance and thermal inertia show a ~50,000-km$^2$ anomaly (Fig. 17). The low rock abundance and low regolith thermal inertia are consistent with a relatively thick deposit of fine-grained, glass-rich materials formed by long-lived Hawaiian-style fire fountaining (Gaddis et al., 1985). A lack of entrained rocks in the eruption plume would explain both its low thermal inertia and rock abundance. Subsequent impacts appear not to have excavated bedrock, implying a thick deposit.

Localized pyroclastic deposits (<2500 km$^2$) are not easily distinguished in the thermal data. Utilizing multiple remote sensing datasets including the Diviner H-parameter, Trang et al. (2017) studied the composition and physical properties of 34 globally-distributed localized pyroclastic deposits, finding that these features do not exhibit a distinct thermal signature. Bennett et al. (2016) studied the pyroclastic deposits within Oppenheimer crater, which were also indistinguishable in the Diviner thermophysical dataset. This could be due to the entrainment of country rock or cap rock during violent Vulcanian eruptions, which are shorter lived, with lower effusion rates than those of the larger regional pyroclastic deposits (Bennett et al., 2016). Alternatively, a similar initial lower-thermal inertia signature at localized deposits may be more quickly erased by impact gardening, owing to their smaller vertical and horizontal extent.

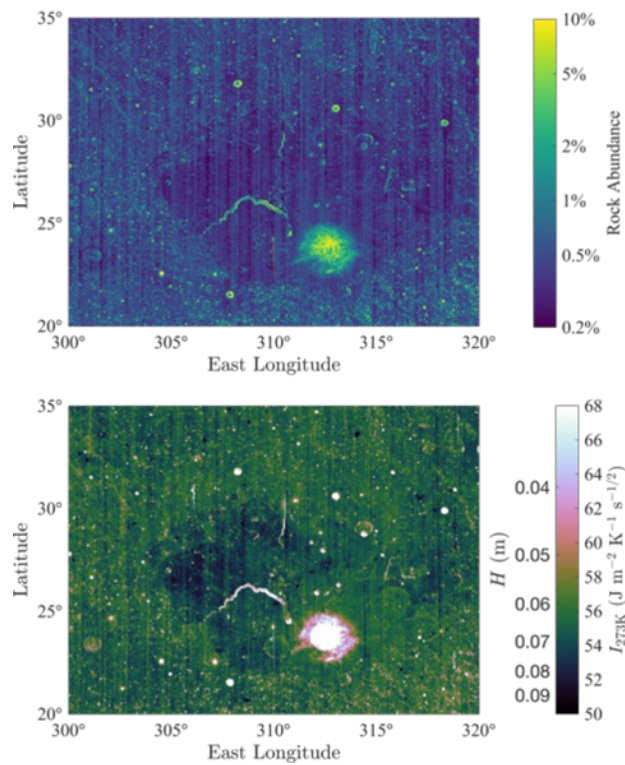

Figure 17: The Aristarchus plateau, showing both low rock abundance (upper panel) and low thermal inertia regolith (lower panel). These features are consistent with a thick deposit of rock-poor materials produced during a long-lived pyroclastic eruption. Also visible in this frame is Aristarchus crater (the near-circular feature with very high thermal inertia centered at 313°E, 23°N) Schröter's Valley, a sinuous rille where mass wasting has apparently exposed higher thermal inertia materials on its steep slopes.

## 5 Discussion

At the global scale (>100 km), Diviner thermal inertia maps show that the lunar regolith is remarkably uniform; variations across the Moon at the 1,000-km scale are < 10%. Lacking a clear hemispheric or maria/highlands dichotomy, we infer a rapid regolith formation process, which homogenizes the upper ~10 cm on timescales < 1 Gyr. In the global view, the regolith thermal inertia maps here correlate loosely with the optical maturity index presented by Lucey et al. (2000). This correlation is also consistent with reworking and overturn of the regolith, leading to finer (hence lower thermal inertia) regolith. A pattern of systematically higher surface conductivity in the older highlands terrains reported by Yu and Fa (2016) does not appear to dominate the best-fit H-parameter, which shows relatively uniform values for maria and highlands units. It is difficult to reconcile this discrepancy, since their factor of >50% enhancement of $K_s$ in the older units would be expected to show up as a systematically higher regolith thermal inertia our model fits. However, we note that aside from the global-scale patterns, most large regional-scale ~30- to 1,000-km features in the $K_s$ maps of Yu and Fa (2016), such as young impact craters and their rays, are also present as high-thermal inertia anomalies in our maps. Perhaps different factors control thermal conductivity at the uppermost surface, in contrast to the bulk layer extending to several centimeters depth. Alternatively, fits to the bolometric temperatures may yield different results due to the presence of surface rocks, which we have explicitly removed. Finally, non-lambertian reflectance behavior may also systematically alter derived thermophysical properties for darker and lighter units. Therefore, more detailed future modeling may seek to better quantify these effects to reveal patterns of vertical stratification at the cm-scale, which may not be captured by our model.

Regional patterns (~10 – 100 km scale) in thermal inertia typically fall into one of three categories: 1) higher thermal inertia anomalies associated with Copernican-aged (< 1 Ga) craters and their ejecta, 2) low thermal inertia anomalies associated with "cold spot" craters younger than ~1 Myr, and 3) low thermal inertia anomalies associated with large regional-scale pyroclastic deposits. We also observe higher thermal inertia materials on slopes where mass wasting has occurred. Rocky ejecta probably account for the higher thermal inertia deposits around Copernican-aged craters, whereas cold spots are likely formed by the in-situ disruption and decompression of the regolith. The difference in the apparent preservation timescales of rocky craters (~1 Gyr) and cold spots (~100 kyr) is likely due to the fact that the rock breakdown process is slower than regolith compaction and overturn; as larger rocks are broken down, their fragments litter the surface even as regolith builds up. Buried rocks within ~10 cm of the surface would also be detectable in the thermal inertia dataset (Elder et al., 2017a), but not the rock abundance dataset. Differences between these two datasets around geologically young craters could be leveraged to estimate a burial timescale.

In the case of the regional pyroclastic deposits, their persistence despite ages >3 Ga in at least some cases (e.g., Head, 1974) indicates a very thick deposit of relatively uniform material. The near absence of rocks at the surface is consistent with similar inferences for the bulk of the pyroclastic deposits using radar techniques (e.g., Carter et al., 2009). Although it is beyond the scope of this work, we suggest that future work could use the Diviner data and LROC images to determine the minimum crater size with rocky ejecta to provide a quantitative thickness estimate for regional pyroclastic deposits. In contrast, the lack of a clear signature in the Diviner data presented here indicates that localized pyroclastic deposits may be explained by the complex Vulcanian eruption model of Bennett et al. (2016). Alternatively, these smaller deposits may be more quickly obscured by rocks and regolith gardening. A low thermal inertia signature was observed at the "irregular mare patch", Ina, as well. On hypothesis that could explain this observation is a pyroclastic eruption (Elder et al., 2017a).

Regolith H-parameter may provide a quantitative tool for estimating the ages of lunar craters, if the trend identified in Section 4.2 holds generally. Such measurements could be combined with estimates from rock abundance (Ghent et al., 2014) for potentially improved accuracy. After roughly 1 Gyr, crater ejecta fade to the background, whereas a lower bound on this possible age-dating technique has not yet been determined. The thermal approach (combining rock abundance and thermal inertia) could be useful for estimating ages of Copernican-aged impact craters where crater density measurements are not possible, for example small or poorly imaged craters. One feature where this technique would be especially interesting is the Tycho antipodal deposit, which shows both elevated rock abundance and thermal inertia. It is not yet clear how these characteristics would originate and evolve through time.

## 6  Conclusions

Lunar surface temperatures measured by Diviner provide a global view of the Moon's regolith thermophysical properties. By using Diviner's multi-spectral brightness temperature measurements, we are able to separate the thermal behavior of the regolith from that of meter-scale and larger rocks. These measurements provide quantitative constraints on the physical properties of the upper ~10 cm of regolith, sampled at 128-ppd (~250-m) horizontal resolution. A vertical profile with an exponential increase in density and conductivity with depth, proposed by Vasavada et al. (2012), fits these data well over the whole Moon at this spatial scale.

We find that the Moon's upper regolith is remarkably uniform at the global scale: an average thermal inertia $I_{273K} \approx 55$ J m$^{-2}$ K$^{-1}$ s$^{-1/2}$ with standard deviation ~2 J m$^{-2}$ K$^{-1}$ s$^{-1/2}$ in the upper 4 – 7 cm. Due to its temperature sensitivity, thermal inertia varies by a factor of ~2 from midnight (35 J m$^{-2}$ K$^{-1}$ s$^{-1/2}$) to noon (70 J m$^{-2}$ K$^{-1}$ s$^{-1/2}$) at the equator. The H-parameter provides a convention for describing the depth-profiles of thermophysical properties, which are temperature-independent. Here, we presented maps of the H-parameter and showed its equivalent fixed-temperature thermal inertia, $I_{273K}$.

No dichotomy in regolith thermal inertia is observed between the maria and highlands. This implies a rapid process of homogenization by impact gardening and lateral mixing within the upper ~10 cm probed by the Diviner measurements. However, significant variations in regolith thermal inertia occur at regional and local scales. Higher thermal inertia materials occur around large impact craters of Copernican age, < 1 Gyr, and also on steep slopes and rilles, where bedrock is exposed by mass wasting. In contrast, low thermal inertia anomalies are predominantly contained in the thousands of "cold spots" – vast disrupted regions surrounding ~1-km and smaller fresh impact craters ~100 kyr in age (Bandfield et al., 2014). Notably, all of these impact-related features are < 1 Gyr old, explaining the general correlation of optical maturity (Lucey et al., 2000) and thermal inertia; features older than about a billion years have been erased by erosion.

Several regional scale (>2500 km$^2$) pyroclastic deposits, such as the Aristarchus plateau, exhibit low thermal inertia indicating thick, rock-poor layers. These features are interpreted to be consistent with long-lived Hawaiian-style fire fountain eruptions. In contrast, smaller local-scale pyroclastic deposits, such as those at Oppenheimer crater (Bennett et al., 2016), do not show a thermal inertia signature, possibly due to a shorter-lived eruption incorporating country rock or cap rock into the deposit.

Impact ejecta show a trend of decreasing thermal inertia with crater age. Enhancements in thermal inertia relative to rock abundance in crater ejecta may be due to a combination of small (< 10 cm) surface rocks and buried larger rocks within the upper ~10 cm. Both the rock abundance and thermal inertia signatures disappear after ~1 Gyr, implying the burial of rocks > 10 cm depth on this timescale. The observed trends suggest a potential age-dating tool, complementary to the technique proposed by Ghent et al. (2014) using rock abundance. This approach would be valid for craters < 1 Ga, and may be useful in situations where crater counts yield unreliable results due to small areas or limited image coverage.

The global, 128-pixel-per-degree maps of regolith H-parameter and thermal inertia presented here will be useful for future investigations of the Moon's geologic history. For example, ray patterns indicate the distribution of ejecta including small rocks in exquisite detail. Could these patterns be used to estimate the contributions of individual impact craters to lunar samples acquired in specific locations? Recent impacts appear to consistently form cold spots, which indicate regolith modification to much greater distances than model predictions (Bandfield et al., 2014); here we quantified the decrease in regolith thermal inertia. By what process does the regolith become fluffier? Near the antipode of Tycho crater, we find concentrations of rocky materials and high-thermal inertia regolith, including ray-like patterns. Future work should investigate the concentration of materials at the antipodes of large impacts on the Moon and other bodies, in order to explain the observed thermophysical anomalies. Improvements to the Diviner thermophysical datasets and models will undoubtedly reveal new and unexpected features. For example, modeling the effects of the full rock size-frequency distribution may better resolve the distinction between rocks and regolith, with implications for selecting safe landing sites for future missions (Elder and Hayne, 2017b). To maximize potential science return from future missions to other planetary bodies, the results shown here underline the need for thermal emission measurements over a range of local times and in multiple spectral channels.

**Acknowledgements:**

The data used in this study are publicly available via the Geosciences Node of the Planetary Data System: http://pds-geosciences.wustl.edu/missions/lro/diviner.htm

Numerical models used in this study are freely available from the authors in C/C++, Python, and MATLAB implementations, and may be used for research purposes with proper citation. Current versions may also be obtained from the GitHub repository: https://github.com/phayne/heat1d

This work was supported by the Lunar Reconnaissance Orbiter project. Part of this work was performed at the Jet Propulsion Laboratory, California Institute of Technology, under contract with the National Aeronautics and Space Administration. The contributions of R. Ghent were funded by a Discovery Grant from the National Science and Engineering Research Council of Canada. P. Hayne gratefully acknowledges support from the Weizmann Institute of Science, while in residence during part of this work. © 2017, all rights reserved.

**Appendix A: Thermal Model**

In this Appendix, we describe the numerical thermal model used to interpret lunar surface temperature data. This finite-difference approach is robust and extensively validated, with heritage from diverse areas of planetary science (e.g., *Morrison* 1969; *Paige*, 1992; *Vasavada et al.*, 1999; *Kieffer*, 2013). The description below is intended to be complete in the sense that it should allow the interested reader to reproduce the model in its entirety. We also present several benchmark tests that can be used as checks on other models.

**A1 Theory**

The variation of temperature $T$ with time $t$ and depth $z$ in a one-dimensional solid medium is governed by the heat equation

$$\rho c_p \frac{\partial T}{\partial t} = \frac{\partial}{\partial z}\left( K \frac{\partial T}{\partial z} \right), \tag{A1}$$

where $\rho$ is the density, $c_p$ is the specific heat, and $K$ is the thermal conductivity of the material. Our numerical model (*Hayne and Aharonson*, 2015) employs a standard finite-difference approximation for the derivatives (*Morrison*, 1969; *Kieffer et al.*, 2013), and has been previously validated using diurnal temperature measurements of the Moon by Diviner (*Hayne et al.,* 2010; *Vasavada et al.*, 2012). The Diviner data are consistent with depth-dependent density, of the form

$$\rho(z) = \rho_d - (\rho_d - \rho_s)e^{-z/H}, \tag{A2}$$

where $\rho_s$ and $\rho_d$ are the bounding densities at the surface and at depths much greater than $H$, which is the scale height of the vertical profile. The thermal conductivity varies with composition, density, and temperature, which can be described by (*Whipple*, 1950):

$$K(T,\rho) = K_c(\rho) + BT^3, \tag{A3}$$

where $K_c$ is the solid phonon conductivity, and $B \sim \sigma \bar{\varepsilon}_0 l$ is the "radiative" conductivity factor, with $\sigma$ the Stefan-Boltzmann constant, $\bar{\varepsilon}_0$ the bolometric infrared emissivity of individual grains, and $l$ the inter-grain spacing. For an inter-grain spacing $l \sim 100$ μm typical of lunar regolith (Carrier et al. 1973), $B \sim 10^{-11}$ W m$^{-1}$ K$^{-4}$. It is convenient to encapsulate the radiative component of the thermal conductivity by the dimensionless parameter $\chi$, through $B = K_c \chi / 350^3$, as in *Mitchell and de Pater* (1994):

$$K = K_c \left[ 1 + \chi \left( \frac{T}{350} \right)^3 \right]. \tag{A4}$$

Based on the experimental data of *Fountain and West* (1970), we assume that the contact conductivity is linearly proportional to density over the relevant temperature range:

$$K_c = K_d - \left( K_d - K_s \right) \frac{\rho_d - \rho}{\rho_d - \rho_s} \tag{A5}$$

where the constants $K_s$ and $K_d$ are the contact conductivity values at the surface and at depth, respectively.

Heat capacity is also temperature dependent; we used the data of *Ledlow et al.* (1992) and *Hemingway et al.* (1981) to derive a polynomial fit

$$c_p(T) = c_0 + c_1 T + c_2 T^2 + \ldots + c_N T^N \tag{A6}$$

with the values of the coefficients $c_i$ given in Table A1, along with other parameter values. The polynomial fit $c_p(T)$ is valid for temperatures from < 90 K to >400 K, but is strictly invalid at temperatures < 1.3 K, where it becomes negative. Additional experimental data are needed in order to better constrain the heat capacity of lunar materials under the extremely low temperatures (< 30 K; Paige et al., 2010b) often encountered at the lunar poles and elsewhere in the solar system.

Table A1: Standard model parameters and physical constants

| Parameter | Symbol | Value | Reference |
|---|---|---|---|
| Solar constant | $S$ | 1361 W m$^{-2}$ | *Kopp and Lean* (2011) |
| Lunar diurnal period (synodic month) | $P$ | 2.55024×10$^6$ s (= 29.5306 d) | *Lang* (2012) |
| Infrared emissivity | $\bar{\varepsilon}$ | 0.95 | *Logan et al.* (1972) and *Bandfield et al.* (2015) |
| Bond albedo at normal solar incidence (lunar average) | $A_0$ | 0.12 | *Vasavada et al.* (2012) |
| Bond albedo at arbitrary solar incidence $\theta$ | $A$ | $A_0 + a\left(\dfrac{\theta}{\pi/4}\right)^3 + b\left(\dfrac{\theta}{\pi/2}\right)^8$ | *Keihm* (1984) |
| Constants | $a$ | 0.06 | This study |
| | $b$ | 0.25 | This study |
| Thermal conductivity | $K$ | $K_c + BT^3$ | *Whipple* (1950) |
| Phonon conductivity | $K_c$ | $K_d - (K_d - K_s)\dfrac{\rho_d - \rho}{\rho_d - \rho_s}$ | *Vasavada et al.* (2012) |
| Surface layer conductivity | $K_s$ | 7.4×10$^{-4}$ W m$^{-1}$ K$^{-1}$ | This study |
| Deep layer conductivity | $K_d$ | 3.4×10$^{-3}$ W m$^{-1}$ K$^{-1}$ | This study |
| Radiative conductivity factor | $B$ | $K_c \chi / (350\,\mathrm{K})^3$ | *Mitchell and de Pater* (1994) |
| Radiative conductivity parameter | $\chi$ | 2.7 | This study and *Vasavada et al.* (2012) |
| Regolith density | $\rho$ | $\rho_d - (\rho_d - \rho_s)e^{-z/H}$ | *Vasavada et al.* (2012) |
| Surface layer density | $\rho_s$ | 1100 kg m$^{-3}$ | *Hayne et al.* (2013) |
| Deep layer density | $\rho_d$ | 1800 kg m$^{-3}$ | *Carrier et al.* (1973) |

| Scale factor | $H$ | 0 to > 0.2 m (avg. = 0.06 m) | This study |
|---|---|---|---|
| Specific heat capacity | $c_p$ | $c_0 + c_1 T + c_2 T^2 + c_3 T^3 + c_4 T^4$ | *Hemingway et al.* (1981), *Ledlow et al.* (1992) |
| Coefficients for specific heat capacity function | $c_0$ | -3.6125 J kg$^{-1}$ K$^{-1}$ | This study |
| | $c_1$ | +2.7431 J kg$^{-1}$ K$^{-2}$ | " |
| | $c_2$ | +2.3616×10$^{-3}$ J kg$^{-1}$ K$^{-3}$ | " |
| | $c_3$ | -1.2340×10$^{-5}$ J kg$^{-1}$ K$^{-4}$ | " |
| | $c_4$ | +8.9093×10$^{-9}$ J kg$^{-1}$ K$^{-5}$ | " |
| Interior heat flow | $Q$ | 0.018 W m$^{-2}$ | *Langseth et al.* (1976) |

### A1.1 Boundary Conditions

At the surface, absorbed insolation and conduction are balanced against infrared emission to space:

$$K \frac{\partial T}{\partial z}\bigg|_{z=0} + Q_s = \bar{\varepsilon}\sigma T_s^4 \tag{A7}$$

Here, the variable $Q_s$ represents the surface energy flux, which is typically equal to the solar heating rate: $Q_s = (1-A)F_\odot$, with incident solar flux $F_\odot$ (measured in watts per square meter). Following the empirical fits of *Keihm et al.* (1984) and *Vasavada et al.* (2012), we adopt an albedo dependent on solar incidence angle, $\theta$,

$$A(\theta) = A_0 + a\left(\frac{\theta}{\pi/4}\right)^3 + b\left(\frac{\theta}{\pi/2}\right)^8 \tag{A8}$$

with slightly updated best-fit constants $a = 0.06$ and $b = 0.25$. On horizontal surfaces, the solar flux is given by (cf. Liou 2002):

$$F_\odot(t) = \begin{cases} \dfrac{S}{R_{AU}^2}\left(\sin\phi\sin\delta + \cos\phi\cos\delta\cos h\right), & \cos h \geq 0 \\ 0, & \cos h < 0 \end{cases} \quad (A9)$$

where $S \approx 1361$ W m$^{-2}$ is the solar constant (*Kopp and Lean*, 2011), $R_{AU}$ is the distance to the Sun in astronomical units, $\phi$ is latitude, $\delta$ is the solar declination angle, and the "hour angle" is $h = 2\pi t/P$, with $P$ the length of the synodic day. It is useful to define a clipping function

$$\psi(x) = \tfrac{1}{2}\left(\cos x + |\cos x|\right) \quad (A10)$$

where vertical bars indicate the absolute value. The insolation function (A9) can then be written in terms of the solar incidence angle $\theta$

$$F_\odot(t) = \frac{S}{R_{AU}^2}\psi(\theta) , \quad (A11)$$

with $\cos\theta = \sin\phi\sin\delta + \cos\phi\cos\delta\cos h$ over the full range of $h$. Finally, to calculate the instantaneous insolation and infrared heating on arbitrary (thermally isolated) slopes, the geometric formulas of *Braun and Mitchell* (1983) and *Aharonson and Schörghofer* (2006) are used.

At the lower boundary $z_*$, the heat flow is dominated by the geothermal flux $Q$, such that the temperature is determined by its gradient:

$$\left.\frac{\partial T}{\partial z}\right|_{z=z_*} = Q/K \quad (A12)$$

**A2  Numerical Solution**

The conducted heat flux, $F = K\,\partial T/\partial z$, is conserved at every depth over a complete annual cycle, for constant orbital elements and in the absence of internal heating (e.g., due to radioactive decay or phase changes). In other words, the time-averaged heat flux at any layer in the model equals the geothermal flux, as long as the model has equilibrated properly. Rewriting the heat equation (A1) in terms of this conserved quantity,

$$\frac{\partial T}{\partial t} = \frac{1}{\rho c_p} \frac{\partial F}{\partial z} \tag{A13}$$

In discrete form, a change in temperature over a time increment $\Delta t$ is given by

$$\Delta T = \frac{\Delta t}{\rho c_p} \frac{\partial F}{\partial z} \tag{A14}$$

The flux across each layer can be approximated by the forward difference formula

$$F_i \approx K_i \frac{T_{i+1} - T_i}{\Delta z_i} \tag{A15}$$

with $\Delta z_i \equiv z_{i+1} - z_i$, and the thermal conductivity $K_i$ is taken to be that between layers $i$ and $i+1$. The flux gradient on the right-hand side of equation (A13) can then be approximated

$$\begin{aligned}
\frac{\partial}{\partial z} F_i &\approx \frac{F_i - F_{i-1}}{\tfrac{1}{2}(\Delta z_i + \Delta z_{i-1})} \\
&\approx \frac{2}{\Delta z_i + \Delta z_{i-1}} \left\{ K_i \frac{T_{i+1} - T_i}{\Delta z_i} - K_{i-1} \frac{T_i - T_{i-1}}{\Delta z_{i-1}} \right\} \\
&\approx \frac{2}{\Delta^3 z_i} \left\{ T_{i-1} K_{i-1} \Delta z_i - T_i (K_{i-1} \Delta z_i + K_i \Delta z_{i-1}) + T_{i+1} K_i \Delta z_{i-1} \right\}
\end{aligned} \tag{A16}$$

where $\Delta^3 z_i \equiv \Delta z_i \Delta z_{i-1} (\Delta z_i + \Delta z_{i-1})$. Equation (A16) can be simplified by defining $\alpha_i \equiv 2 K_{i-1} \Delta z_i / \Delta^3 z_i$, and $\beta_i \equiv 2 K_i \Delta z_{i-1} / \Delta^3 z_i$, such that combined with equation (A14), the temperature at each layer $i$ is updated over a time increment $\Delta t$ using

$$T_i^{(n+1)} = T_i^{(n)} + \frac{\Delta t}{(\rho c_p)_i} \left\{ \alpha_i T_{i-1}^{(n)} - (\alpha_i + \beta_i) T_i^{(n)} + \beta_i T_{i+1}^{(n)} \right\} \tag{A17}$$

where $n$ is the previous time step. In practice, the numerical grid remains fixed throughout each simulation, such that the factors

$$\begin{aligned}
p_i &\equiv 2\Delta z_i / \Delta^3 z_i \\
q_i &\equiv 2\Delta z_{i-1} / \Delta^3 z_i
\end{aligned} \tag{A18}$$

only need to be calculated once, and the coefficients in equation (A17) are calculated at each time step using $\alpha_i(t) = p_i K_{i-1}(t)$ and $\beta_i(t) = q_i K_i(t)$. Even though the density is typically constant in time, the heat capacity varies with temperature (Eq. (A6)), so the pre-factor $\Delta t / \rho c_p$ must be calculated at each layer, at each time step.

For numerical stability, the one-dimensional Fourier mesh number must be less than 0.5:

$$\text{FoM}_1 = \frac{\Delta t}{\rho c_p}(\alpha + \beta) < 0.5 \tag{A19}$$

such that the maximum time step is

$$\Delta t_{max} = \left[\frac{\rho c_p}{2(\alpha + \beta)}\right]_{min} \sim \Delta z^2 \tag{A20}$$

where the subscript "min" refers to the minimum value among all layers.

### A2.1 Numerical solution of the boundary conditions

At each time step, the upper boundary condition (A7) is solved using Newton's root-finding method to iteratively improve the estimate for the surface temperature $T_0$:

$$T_0' = T_0 + \delta T \tag{A21}$$

$$\delta T = -f/f' \tag{A22}$$

where $f$ is the function whose zeros we seek:

$$f \equiv \bar{\varepsilon}\sigma T^4 - K\frac{\partial T}{\partial z} - Q_s = 0. \tag{A23}$$

For improved accuracy, here we approximate the spatial derivative using a three-point numerical scheme,

$$\frac{\partial T}{\partial z} \approx \frac{-3T_0 + 4T_1 - T_2}{2\Delta z_0} \tag{A24}$$

where subscripts refer to model layer indices. Then the boundary condition is approximated

$$f \approx \bar{\varepsilon}\sigma T_0^4 - Q_s - \left(K_{c,0} + B_0 T_0^3\right)\left[\frac{-3T_0 + 4T_1 - T_2}{2\Delta z_0}\right]. \tag{A25}$$

The partial derivative with respect to temperature is

$$\begin{aligned}f' &= \frac{\partial f}{\partial T} \\ &= 4\bar{\varepsilon}\sigma T^3 - \frac{\partial}{\partial T}\left[K\frac{\partial T}{\partial z}\right] \\ &= 4\bar{\varepsilon}\sigma T^3 - \frac{\partial K}{\partial T}\frac{\partial T}{\partial z} - K\frac{\partial}{\partial T}\frac{\partial T}{\partial z}\end{aligned} \tag{A26}$$

Numerical approximations of the latter two terms are given by

$$\frac{\partial K}{\partial T}\frac{\partial T}{\partial z} \approx 3B_0 T_0^2 \left[\frac{-3T_0 + 4T_1 - T_2}{2\Delta z_0}\right] \tag{A27}$$

and

$$\frac{\partial}{\partial T}\frac{\partial T}{\partial z} \approx \frac{\partial}{\partial T_0}\left[\frac{-3T_0 + 4T_1 - T_2}{2\Delta z_0}\right] = -\frac{3}{2\Delta z_0}. \tag{A28}$$

Then the first derivative of the boundary condition is approximated by

$$f' \approx 4\bar{\varepsilon}\sigma T_0^3 - 3B_0 T_0^2 \left[\frac{4T_1 - 3T_0 - T_2}{2\Delta z_0}\right] + \frac{3}{2\Delta z_0}\left(K_{c,0} + B_0 T_0^3\right) \tag{A29}$$

Using equations (A25) and (A29), equation (A21) is solved iteratively, until $\delta T \ll 1$ K.

At the bottom boundary $N$, the finite difference form of equation (A12) is straightforward:

$$T_N = T_{N-1} + \frac{Q}{K_{N-1}}\Delta z_{N-1} \tag{A30}$$

### A2.2 Skin depth, layer thicknesses, and model domain

Thermal skin depth is a useful quantity that describes the depth of penetration of a periodic temperature wave, such as the diurnal cycle (Carslaw and Jaeger, 1959):

$$z_s = \sqrt{\frac{\kappa P}{\pi}} \tag{A31}$$

where $\kappa = K/\rho c_p$ is the thermal diffusivity. A surface temperature oscillation with amplitude $\Delta T_0$ is attenuated by $\Delta T_i \approx \Delta T_0 e^{-z_i/z_s}$. Before each model run, $z_s$ is calculated based on the expected temperature range (since $\kappa$ is temperature-dependent), and the vertical grid is defined using the scheme:

$$\begin{aligned} z_0 &= 0 \\ z_i &= z_{i-1} + \Delta z_{i-1}, \quad i = 1,...,N \end{aligned} \tag{A32}$$

with geometrically-increasing grid spacing

$$\begin{aligned} \Delta z_0 &= z_s/m \\ \Delta z_i &= \Delta z_{i-1}\left(1 + \frac{1}{n}\right), \quad i = 1,...,N \end{aligned} \tag{A33}$$

The factors $m$ and $n$ can be adjusted for accuracy or speed, and we found optimal values to be close to $m = 10$ and $n = 5$ for most modeling work (cf. *Kieffer*, 2013). For the global H-parameter fitting presented here, we used $m = 10$ and $n = 20$ for improved accuracy, i.e. all depths within < 1 K of the asymptotic values for $m, n \to \infty$. The depth of the lowermost layer is set to be a multiple of the skin depth, typically ~10, in which case the diurnal wave is damped by a factor $\exp(-10) \approx 10^{-5}$ at the bottom boundary. With these parameters, a typical model run uses a grid with ~15 – 20 layers.

**A2.3 Temperature initialization**

Equilibration times can be greatly reduced if an appropriate initial temperature profile is chosen. At the surface, we choose an initial temperature representing instantaneous radiative equilibrium at noon:

$$T_0 = \left(\frac{1-A}{\bar{\varepsilon}\sigma} F_\odot\right)^{1/4} \tag{A34}$$

At the lower boundary, the temperature is initialized to the equilibrium mean annual temperature for an isothermal body,

$$T_N = T_0/\sqrt{2} \tag{A35}$$

All other layers are initialized using the formula

$$T_i = T_N - (T_N - T_0)e^{-z_i/H} \tag{A36}$$

which results in a more uniform temperature profile for larger values of $H$, consistent with higher thermal inertia materials.

### A2.4 Equilibration and computational cost

Equilibration times are defined such that the year-to-year variation in instantaneous temperature at any depth is << 1 K. Experiments over the full parameter space have shown that this level of repeatability is established after ~2 yr, but we conservatively require typically 5 yr equilibration before reporting results. This period allows both seasonal and annual oscillations to be completely damped, as indicated by inter-annual temperature deviations. Computational costs are low, with each 5-yr simulation requiring ~100 milliseconds of processor time on a single 2.8-GHz Intel Core i7.

### A3 Model validation and data fitting

Lunar surface and subsurface temperatures from remote sensing and in-situ experiments were used to validate the model. These include heat flow probe measurements from Apollo 15 and 17 (*Keihm et al.*, 1973a,b), which provide both surface and deep subsurface (~1 m) temperature measurements during the day and night. We also leveraged the extensive Diviner dataset to constrain the model at all latitudes and local times. In particular, *Keihm et al.* (1973a,b) derived diurnal

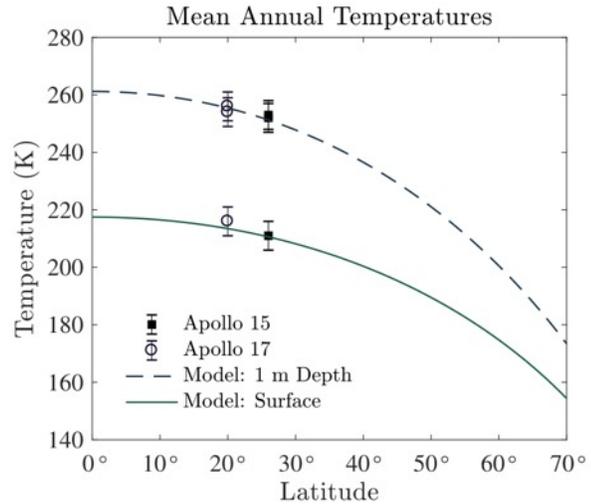

**Figure A1**: Diurnal average temperatures measured at the Apollo 15 & 17 landing sites [*Keihm et al.*, 1973a,b] compared to model results. All model parameters are mean surface temperatures of 216 ±5 K and 211 ±5 K, and diurnal mean subsurface temperatures of 256 ±5 K (1.3 m depth) and 252 ±5 K (0.8 m depth) at the Apollo 17 (20°N

latitude) and 15 (26°N latitude), respectively. Using Diviner data, *Vasavada et al.* (2012) found typical equatorial temperatures of ~385 K at local noon, 101 K at midnight, and 95 K just before sunrise. These and other constraints are summarized in Table A2. Physical properties of the lunar regolith, including thermal conductivity, density, and heat capacity, have been constrained by a variety of methods (*Carrier et al.*, 1991). In Table A2, we have also compiled some of these properties, which show general agreement with the values derived from the Diviner data. Discrepancies can largely be attributed to the different techniques used, particularly the unavoidable changes that occurred to regolith samples during acquisition and handling in the case of in situ and laboratory analyses.

We found that in order to simultaneously fit the Apollo heat flow probe surface and subsurface temperatures, it was necessary to adjust the thermal conductivity from *Vasavada et al.* (2012) slightly, to $K_s = 7.4 \times 10^{-4}$ W m$^{-1}$ K$^{-1}$ and $K_d = 3.4 \times 10^{-3}$ W m$^{-1}$ K$^{-1}$. This is understood to be the result of the formulation of the radiative conductivity (equation (A4)) as proportional to $K_c(\rho)$ rather than the constant $K_s$. Based on spectroscopic studies of lunar materials (*Donaldson Hanna et al.*, 2012), we adjusted the bolometric infrared emissivity downward from 0.98 to 0.95, which partly offsets the lower thermal conductivity with respect to nighttime temperatures. Finally, eclipse cooling measured by Diviner (*Hayne et al.*, 2013), along with data from Apollo core samples, yields an improved estimate of the uppermost regolith density of ~1100 kg m$^{-3}$. Figures A1-A4 show results of typical model runs using the standard parameter set.

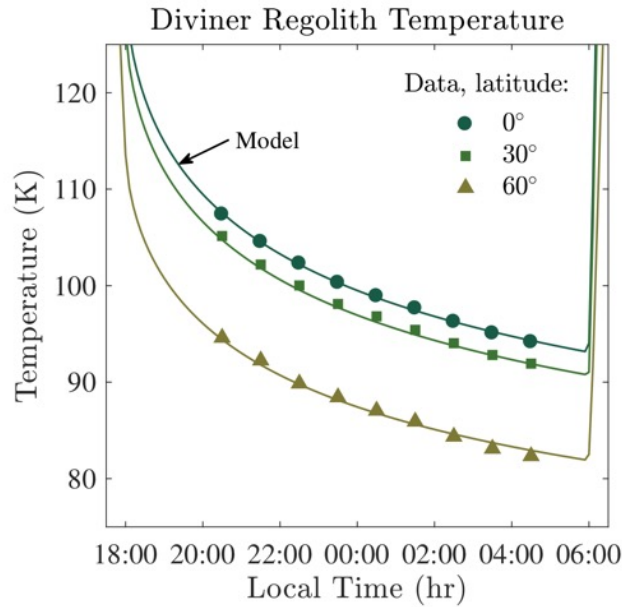

Figure A2: Example model fit to the nighttime Diviner equatorial (±1° latitude) regolith temperature data, selected from the 128-ppd gridded dataset. The model is a result using fixed $H = 0.06$ m, $\rho_s = 1100$ kg m$^{-3}$, and $\rho_d = 1800$ kg m$^{-3}$, to derive best fit values $K_s = 7.4 \times 10^{-4}$ W m$^{-1}$ K$^{-1}$, and $K_d = 3.4 \times 10^{-3}$ W m$^{-1}$ K$^{-1}$.

Table A2: Lunar thermal modeling constraints

| Parameter | Value | Depth | Latitude | Reference |
|---|---|---|---|---|
| Diurnal mean temperature | 216 (±5 K) | Surface | 20°N | *Keihm et al.* (1973a) |
| " | 256 K (±5 K) | 130 cm | 20°N | " |
| " | 211 K (±5 K) | Surface | 26°N | *Keihm et al.* (1973b) |
| " | 252 K (±5 K) | 83 cm | 26°N | " |
| Peak noontime temperature at equator ($A_0 = 0.12$) | 385 K | Surface | Equator | *Vasavada et al.* (2012) |
| Midnight temperature at equator ($A_0 = 0.12$) | 101 K | Surface | Equator | " |
| Minimum nighttime temperature ($A_0 = 0.12$) | 95 K | Surface | Equator | " |
| Density | 1100 kg m$^{-3}$ | ~0 cm | 26°N | *Carrier et al.* (1973) |
| " | 1600 kg m$^{-3}$ | 0-30 cm | 26°N | " |
| " | 1800-1900 kg m$^{-3}$ | 30-60 cm | 26°N | " |
| Thermal conductivity | 0.9-1.5 x10$^{-3}$ W m$^{-1}$ K$^{-1}$ | 0-2 cm | 20°N | *Keihm et al.* (1973a,b) |
| " | 0.9-1.3 x10$^{-2}$ W m$^{-1}$ K$^{-1}$ | > 50 cm | 20-26°N | *Langseth et al.* (1976) |
| " | 0.6 x10$^{-3}$ W m$^{-1}$ K$^{-1}$ | < 10 cm | Equatorial | *Jones et al.* (1975) |
| " | 0.6 x10$^{-3}$ W m$^{-1}$ K$^{-1}$ | ~0 cm | Equatorial | *Vasavada et al.* (2012) |
| " | 7.0 x10$^{-3}$ W m$^{-1}$ K$^{-1}$ | ~1 m | Equatorial | " |
| Thermal diffusivity, $K/(\rho c_p)$ | 0.4-1.0 x10$^{-8}$ m$^2$ s$^{-1}$ | 0-2 m | 20-26°N | Langseth et al. (1976) |

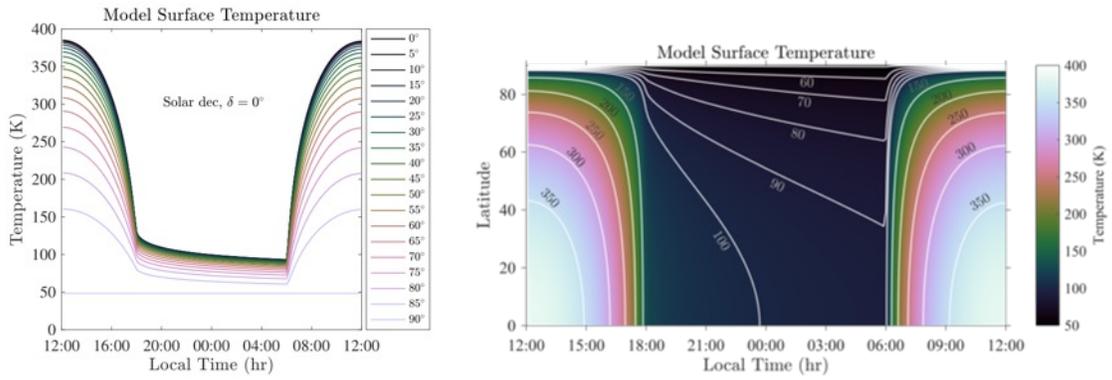

Figure A3: Model surface temperature curves at different latitudes, using the standard parameters from Table A1, with $H$ = 0.06 m. Contours on the right panel are plotted every 10 K at night and every 50 K during the day.

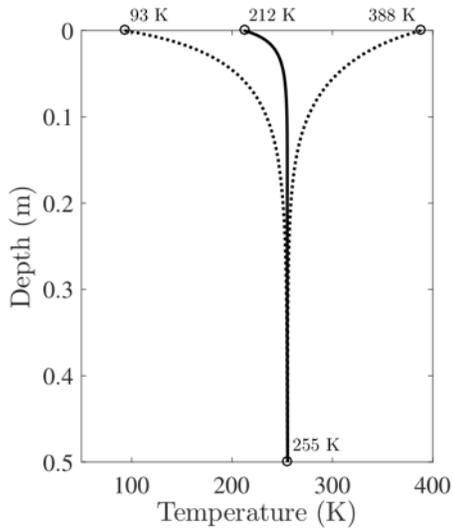

Figure A4: Minimum, average, and maximum temperature profiles for the standard lunar thermal model.

**Appendix B: Heating of Regolith by Rocks**

Rocks remain warmer than regolith during the lunar night, due to their higher thermal inertia. Radiation and conduction from rocks may increase regolith temperatures in rocky areas (Davidsson and Rickman, 2014), affecting derived quantities such as regolith thermal inertia or the H-parameter. Mutual heating by exchange of radiation between warm and cold surfaces is also known to be important on asteroids (e.g., Delbo et al., 2017). Here, we use simple analytic and numerical models of rock-regolith heating on the Moon, to quantify its effects on surface temperatures and measured thermal emission.

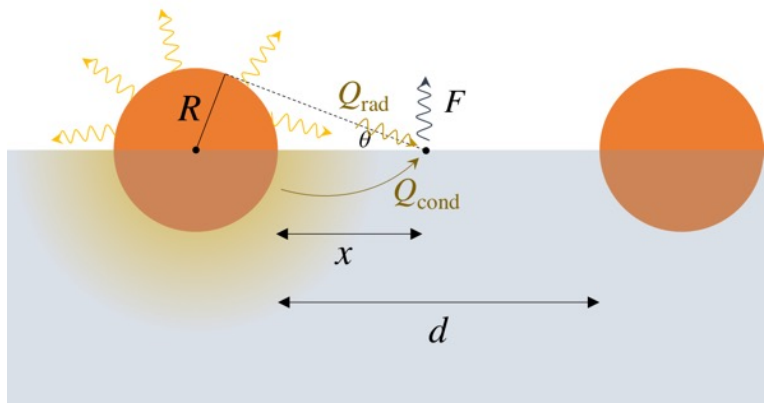

Figure B1: Schematic representation of regolith energy balance near warm rocks, which are modeled as spherical bodies halfway buried in regolith. The distance from the edge of the rock is $x$, and the distance between rocks is $d$.

**B1 Energy balance**

Regolith surface temperatures $T_{\text{reg}}$ at night are determined by the balance of heating by conduction, $Q_{\text{cond}}$, and infrared radiation, $Q_{\text{rad}}$, against cooling by thermal emission to space, $F = \varepsilon \sigma T_{\text{reg}}^4$:

$$Q_{\text{cond}} + Q_{\text{rad}} - F = 0 \tag{B1}$$

The energy balance for a regolith element is depicted in Figure B1. Heat conduction from a nearby rock a distance $x$ away with temperature $T_{\text{rock}}$ is given by

$$Q_{\text{cond}} = K \frac{T_{\text{rock}} - T_{\text{reg}}}{x} \tag{B2}$$

where $K$ is the thermal conductivity of the regolith. Radiant heating of the regolith by a hemispherical rock of radius $R$ is

$$Q_{\text{rad}} = \frac{\varepsilon \sigma T_{\text{rock}}^4}{2\pi} \Omega (1 - A_{\text{IR}}) \tag{B3}$$

where the solid angle is $\Omega = \pi(1 - \cos\theta)$, and the angle $\theta$ subtended by the rock is given by $\sin\theta = R(R^2 + x^2)^{-1/2}$. The fraction of infrared radiation absorbed is given by Kirchhoff's law, $(1 - A_{\text{IR}}) = \varepsilon$. Combining these expressions, we have

$$Q_{\text{rad}} = \frac{1}{2}\varepsilon^2 \sigma T_{\text{rock}}^4 (1 - \cos\theta) \tag{B4}$$

Figure B2 shows an example of the relative contributions of conduction and infrared radiation to heating regolith near a 1-m rock. Here, we assumed a typical nighttime rock temperature $T_{\text{rock}} = 225$ K, background regolith temperature $T_{\text{reg}} = 109$ K, and regolith thermal conductivity $K = 2.5 \times 10^{-3}$ W m$^{-1}$ K$^{-1}$. Conduction dominates heating rates very close to the rock, dropping to negligible values at ~2 cm ($x/R = 0.02$), whereas radiant heating decays more gradually and is important to distances of ~0.8 m ($x/R = 0.8$).

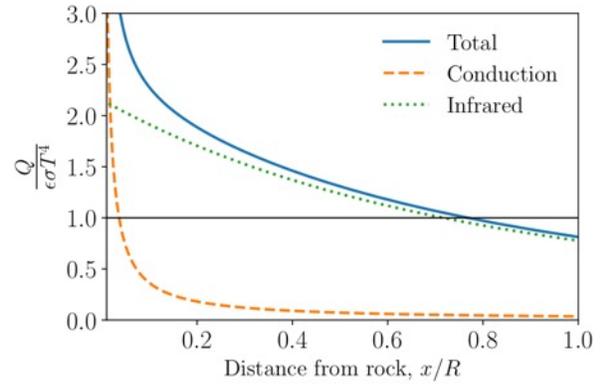

Figure B2: Relative contributions to regolith heating by conduction and radiation from a nearby rock of radius $R$ at distance $x$. The horizontal line at a value of 1.0 indicates where the heating rates are equal to the rate of heat loss by thermal emission.

## B2  Effects of rock-heating on regolith surface temperatures

### B1.1  Analytic approximation

Considering a small change in temperature $\Delta T$ due to rock-regolith heating, we can write equation (B1) as

$$K \frac{T_{\text{rock}} - (T_{\text{reg}} + \Delta T)}{x} + Q_{\text{rad}} = \varepsilon \sigma (T_{\text{reg}} + \Delta T)^4 \tag{B5}$$

Since $\Delta T \ll T_{\text{reg}}$, we neglect terms in $\Delta T^2$ and higher order, to make the approximation

$$\left(T_\text{reg} + \Delta T\right)^4 \approx T_\text{reg}^4 + 4T_\text{reg}^3 \Delta T + \cdots \qquad (B6)$$

Then rearrange and solve

$$K\frac{T_\text{rock} - T_\text{reg}}{x} - \varepsilon\sigma T_\text{reg}^4 + Q_\text{rad} \approx \left(4\varepsilon\sigma T_\text{reg}^3 + K/x\right)\Delta T \qquad (B7)$$

such that

$$\Delta T(x) \approx \frac{Q_\text{cond} + Q_\text{rad} - F}{4\varepsilon\sigma T_\text{reg}^3 + \frac{K}{x}} \qquad (B8)$$

To determine the maximum distance to which a rock affects regolith temperatures, we set $\Delta T = 0$, and solve numerically for $x = x_\text{max}$.

### B1.2  3-d numerical calculations

As a check on the 1-d analytic approach, we also performed 3-d numerical calculations of regolith temperatures near rocks. For this set of simulations, we used the COMSOL Multiphysics package, which has been successfully used for similar heat transfer problems for planetary science (e.g., Piqueux and Christensen, 2009). The COMSOL mesh (Fig. B3) simulates a spherical rock embedded in regolith, such that a hemisphere protrudes upward, with thermophysical properties consistent with those of Bandfield et al. (2011). The model domain is 10 m × 10 m × 1 m. Each simulation includes the full diurnal cycle, with explicit accounting for direct and indirect solar and infrared radiation. For comparison with the analytic approximation, we measured a surface temperature profile extending outward from the edge, perpendicular to the late-afternoon shadow. Results show that the analytic approximation (equation B8) accurately reproduces the more realistic 3-d simulation in this case (Fig. B3). We therefore have confidence in applying the analytic model for $\Delta T(x)$ more generally to estimate the effects of rock-regolith heating on surface temperatures, which are presented below.

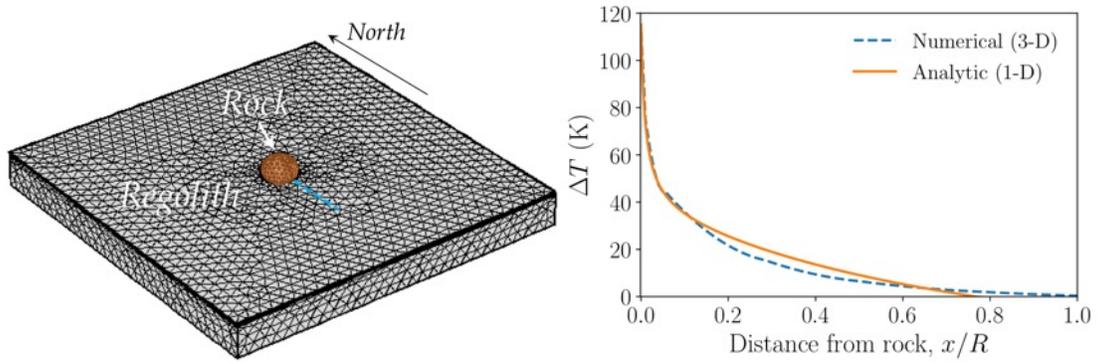

Figure B3: Temperatures at the regolith surface near a warm rock at night. The left panel shows an example computational mesh for the 3-d model, with a spherical rock embedded halfway into the regolith. Profiles (blue line) are taken away from the shadow direction (which is E-W for the equatorial case here). The right panel shows temperature difference $\Delta T$ calculated using both the numerical 3-d model and the analytic 1-d model. In both cases, heating > 1 K extends to a distance $\sim 0.8R$.

### B3 Typical spacing of rocks and extent of regolith warming

Given a rock abundance (i.e., area fraction) $C$ and rock diameter $D = 2R$, the mean distance between rocks is

$$d \sim \left(\frac{A_{\text{rock}}}{C}\right)^{1/2} - D = \left(\frac{1}{\sqrt{C}} - 1\right)D \tag{B9}$$

where $A_{\text{rock}} \approx D^2$ is the rock area. Average rock concentrations on the lunar surface are $C \approx 0.4\%$ (Bandfield et al., 2011). At this concentration, meter-sized rocks are spaced by $d \approx 30$ m, and the average distance to a rock of this size on the lunar surface is roughly $d/4 \approx 7.5$ m. The fractional area of regolith heated by radiation and conduction from rocks is $(\delta A/A_{\text{rock}})C = [(x_{\max}/R)^2 + 2x_{\max}/R]C$, where $x_{\max}$ is the maximum lateral extent of rock-regolith warming described above. (Note that $x_{\max} < d/2$, which is half the typical distance between rocks.) Defining a scaled distance $x' = x/R$, the total fractional area not affected by rocks or rock heating is

$$\alpha = 1 - C(1 + x'_{\max}) \tag{B10}$$

### B4 Effects of regolith-heating by rocks on Diviner measurements

In the simplified two-component rock/regolith model, the measured spectral radiance, $I_\lambda$ (SI units: W m$^{-2}$ sr$^{-1}$ m$^{-1}$) at a specific wavelength $\lambda$ is

$$I_\lambda = \varepsilon_\lambda C B_\lambda(T_{\text{rock}}) + \varepsilon_\lambda(1 - C)B_\lambda(T_{\text{reg}}) \tag{B11}$$

where $B_\lambda(T)$ is the Planck function, and $\varepsilon_\lambda$ is the spectral emissivity of the surface. Heating by rocks increases thermal emission from the surrounding regolith, with each rock's sphere of influence extending to a distance $x_{max}$. Measured brightness temperatures are therefore higher than would be predicted by the simplified two-component model. Including the effects of rock-heating on regolith temperatures, the measured radiance is the sum of contributions from the rocks, the heated regolith region, and the background un-heated regolith:

$$I_\lambda = I_{rock} + I_{reg,heated} + I_{reg} \tag{B12}$$

$$= \varepsilon_\lambda C B_\lambda(T_{rock}) + \varepsilon_\lambda \frac{C}{A_{rock}} \int_0^{x_{max}} B_\lambda\left(T_{reg} + \Delta T\right) 2\pi(R+x)dx + \varepsilon_\lambda \alpha B_\lambda(T_{reg})$$

Brightness temperatures can be calculated from the spectral radiance using the inverse of the Planck function, $B_\lambda^{-1}(I_\lambda)$. Figure B4 shows example brightness temperature curves for this three-component model, and their differences from the simplified two-component model. For typical rock abundance values $C < 1\%$, the error in brightness temperatures predicted by the simplified model are < 1 K.

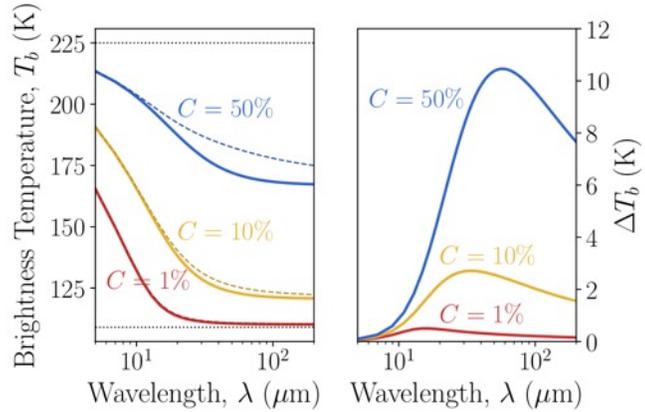

Figure B4: Simulated brightness temperatures for three rock abundance values: $C = 1\%$, 10%, and 50%. The left panel shows results for the simplified two-component model (solid curves) and the three-component model including rock-heating (dashed curves). The right panel shows the differences between these two models.

**B1.3 Diviner brightness temperatures**

Diviner measures radiance across finite spectral bands, with response functions $f_j(\lambda)$ for each spectral channel, $j$. Regolith heating by rocks affects each channel differently, with measured radiance (W m$^{-2}$ sr$^{-1}$)

$$I_j = \int_0^\infty f_j(\lambda) I_\lambda d\lambda \tag{B13}$$

Brightness temperatures $T_j$ are determined from a given radiance distribution by interpolation with a look-up table calculated by replacing $I_\lambda$ with the Planck radiance $B_\lambda(T_j)$.

Figure B5 shows modeled Diviner brightness temperatures in the three primary channels (6, 7, 8) used to derive rock abundance and regolith temperature. We find that for rock concentrations < 3%, all three channels show temperature increases of < 1 K for the rock-heating model relative to the standard two-component model. When rock abundance rises above ~10%, warming of the regolith by rocks becomes significant, and must be modeled explicitly in deriving thermophysical properties.

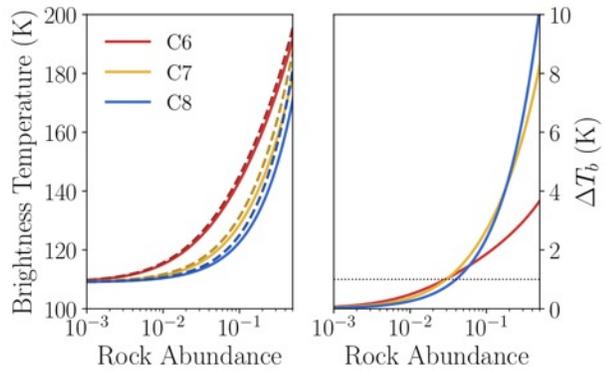

Figure B5: Modeled brightness temperatures in Diviner channels 6, 7, and 8, over a range of rock abundance values. In the left panel, the standard two-component model (solid curves) is compared to the more realistic model including regolith heating by rocks (dashed curves). The right panel shows their difference ($\Delta T_b$). The horizontal dotted line indicates $\Delta T_b = 1$ K, which is a typical noise level for nighttime temperature measurements; below this line, rocks do not measurably affect derived regolith thermophysical properties